\documentclass[11pt]{JHEP3}

\usepackage{epsfig,graphicx,amsfonts,amsthm,amsmath,epsf,amssymb,latexsym,cite,enumerate,shadow,array}
\usepackage[all,ps,matrix,arrow,import,arc]{xy}
\xyoption{matrix}\xyoption{arrow}

\newtheorem{thm}{Theorem}[section]        \newtheorem{lemma}[thm]{Lemma}	
  	\newtheorem{prop}[thm]{Proposition}
  	\theoremstyle{definition}

\DeclareFontFamily{U}{rsf}{} \DeclareFontShape{U}{rsf}{m}{n}{  <5> <6> rsfs5 <7> <8> <9> rsfs7 <10-> rsfs10}{}
\DeclareMathAlphabet\Scr{U}{rsf}{m}{n} \DeclareMathAlphabet\mathbi{U}{cmr}{bx}{it}

\def\CY{Calabi-Yau}	
\def\roof{\mbox{\tiny \mbox{$\!\vee$}}}	\def\comp{\mbox{\scriptsize \mbox{$\circ \,$}}}

\def\O{\mathcal{O}} \def\c#1{\mathcal{#1}}
\def\C{{\mathbb C}}\def\P{{\mathbb P}} \def\Q{{\mathbb Q}}\def\F{{\mathbb F}}
 \def\Z{{\mathbb Z}}		\def\N{{\mathbb N}}
\def\D{\mathbf{D}}	 
\def\iso{\cong}
\def\H{\operatorname{H}}

\def\Hom{\operatorname{Hom}} 	
\def\Ext{\operatorname{Ext}}  
\newcommand{\mytext}[1]{\text{\upshape#1}}   

\def\Spec{\operatorname{Spec}}

\def\rk{\operatorname{rk}}	\def\dim{\operatorname{dim}}

\def\ch{\operatorname{\mathrm{ch}}}

\def\ms#1{\mathsf{#1}}

\def\ses#1#2#3{\xymatrix@1{0 \ar[r] & #1 \ar[r] & #2 \ar[r] & #3 \ar[r] & 0}}

\def\cale{\mathcal{E}}

\def\calb{\mathcal{B}}
\def\calo{\mathcal{O}}

\def\Ext{\mbox{Ext}}
\def\Hom{\mbox{Hom}}

\def\ch{\mbox{ch}}
\def\deg{\mbox{deg}}
\def\Td{\mbox{Td}}

\def\be{\begin{equation}}
\def\ee{\end{equation}}

\title{Exceptional collections and D-branes probing toric singularities}
\author{Christopher P. Herzog\\
Department of Physics\\
University of Washington \\ 
Seattle, WA  98195-1560 USA\\
\email{herzog@phys.washington.edu}}
\author{Robert L. Karp	\\
Department of Physics	\\ 
Rutgers University  	\\
Piscataway, NJ 08854-8019 USA	\\
\email{karp@physics.rutgers.edu}}

\preprint{hep-th/0507175\\
NSF-KITP-05-52}

\abstract{
We demonstrate that a strongly exceptional collection on a singular toric surface can be used to derive the gauge theory on a stack of D3-branes probing the Calabi-Yau singularity caused by the surface shrinking to zero size.  A strongly exceptional collection, i.e., an ordered set of sheaves satisfying special mapping properties, gives a convenient basis of D-branes. We find such collections and analyze the gauge theories for weighted projective spaces, and many of the $Y^{p,q}$ and $L^{p,q,r}$ spaces. In particular, we prove the strong exceptionality for all $p$ in the $Y^{p,p-1}$ case, and similarly for the $Y^{p,p-2r}$ case.
}
\begin{document}

\section{Introduction}

Determining the low-energy gauge theory on a stack of D-branes probing a 
Calabi-Yau singularity is an important, interesting, and in general unsolved problem.
These D-brane constructions can be used to build flux vacua in string theory, and they
play an important role in the AdS/CFT correspondence, where they yield
 a geometric understanding of strongly coupled gauge theories.  While many
simple singularities, such as the conifold and orbifolds (see for example 
\cite{Klebanov-Witten} and \cite{DM}), have been successfully
analyzed, the general case remains elusive.

We analyze D-branes probing the singularity of toric \CY\ three-folds. More precisely, our spaces are local \CY\ varieties, and can be represented as the total space of the canonical sheaf of a singular compact complex surface. While not all Calabi-Yau singularities are toric, the toric subset provides a vast number of interesting examples which, because of their three $U(1)$
isometries, are easier to understand. The conifold, ${\mathbb C}^3/{\mathbb Z}_n$ and ${\mathbb C}^3/{\mathbb Z}_n\times {\mathbb Z}_n$, where the orbifold action preserves the $SU(3)$
holonomy, are all examples of toric Calabi-Yau singularities. 

This paper in large part is motivated by the discovery of  metrics on an infinite class of toric Calabi-Yau three-fold cones  
\cite{Gauntlett, Cvetic} -- the $Y^{p,q}$ and $L^{p,q,r}$ spaces -- and subsequent developments.  
Let $Y$ denote the five dimensional level surface of the cone $X$.  This class of metrics describes all toric
Calabi-Yau singularities where the topology of $Y$ is $S^2\times S^3$. Previously only the metrics for the conifold and ${\mathbb C}^3$ were known!  The new explicit
metrics allow for independent confirmation of the results that we obtain using more abstract algebro-geometric techniques.

We study these Calabi-Yau singularities using exceptional collections \cite{Rudakov}.
Exceptional collections are in essence a ``nice'' basis of D-branes
for describing the gauge theory.  More precisely, an exceptional 
collection is an ordered set of coherent sheaves satisfying special
requirements regarding the maps between the sheaves.  These collections do not
actually live on the Calabi-Yau cone but rather on the base of the cone. Here again the \CY\ space is viewed as 
the canonical sheaf $K_V$ over a given compact surface $V$.

Exceptional collections have appeared in the physics literature before.
Applied first to Landau Ginzburg models \cite{zaslow, hiv, mayr}, they
more recently have reappeared in the AdS/CFT literature \cite{unify, Wijnholt}. Their connection with monodromy was elucidated in \cite{Aspinwall:2001zq}.

Exceptional collections are a useful tool for studying CY singularities for a number
of reasons:
\begin{enumerate}
\item 
Although we study only toric examples in this paper, 
exceptional collections can be used to study
non-toric singularities as well, as pointed out in the 
AdS/CFT context in \cite{unify}.  
For example, collections exist for all del Pezzo singularities, only
some of which are toric \cite{Karpov:Nogin, Wijnholt, HerzogWalcher}.
Hopefully, insights gained here
can be applied to the general case.  
\item
Exceptional collections 
provide a rigorous and constructive way of converting geometry into gauge theory.
Having constructed the gauge theory from the exceptional collection, one can
be {\em absolutely sure} the gauge theory is the low energy effective description of 
the D-branes probing the singularity.  
\item
These
collections allow one to understand all the different Seiberg dual phases of the
gauge theory.  To each singularity, one can typically associate many gauge theories,
all related to each other by Seiberg duality \cite{Seiberg}.  In the exceptional collection language,
Seiberg duality is realized as a special sequence of mutations. 
On the other hand, methods adapted to toric geometry allow one to
access only the ``toric" phases, where the ranks of the gauge groups are all equal \cite{FengHananyHe, BeasleyPlesser}.
\item
Algorithms that convert an exceptional collection into a quiver can be implemented on 
a computer and are typically quite fast. With the computer technology available today, these computations generically take seconds to run. Even the more complicated examples take only a couple of minutes.
\end{enumerate}

This paper is first and foremost a proof of principle that exceptional collections exist
and are useful for studying singularities where a singular two dimensional complex 
surface $V$ shrinks to zero size. More specifically, the singular \CY\ variety $X$ arises as a particular limit
point in the K\"ahler moduli space of the canonical bundle over $V$, where  $V$ itself undergoes a generalized Type II divisorial contraction.

Previously in the physics literature  exceptional collections were used only for smooth surfaces $V$, which essentially meant that $V$ was del Pezzo. All of our interesting examples take $V$ to be a singular toric surface. 
We will find in some cases more convenient to think of $V$ as a stack, rather than
just as a space.  In these cases, we need the stackiness in order to keep track of the extra  information
about how $V$ is embedded in the cone.\footnote{Although $V$ is singular, the cone will generically be smooth everywhere
except at the ``tip''.}
Although $V$ can only have orbifold singularities,\footnote{Toric varieties are always normal, and hence, in two dimensions, the singularities are isolated, and in fact quotient.} 
shrinking $V$ typically produces more general types of singularities in the \CY\ than orbifolds.

In deriving the quiver from the singularity, the first step is to discover an exceptional
collection.  While we do not have a general 
algorithm for producing the collection, we do have several useful tools, which we can combine:
\begin{itemize}
\item We know how to write down an appropriate collection for any weighted projective space ${\mathbb P}^2(a,b,c)$.
\item Given a surface and a collection on it, we can generate a collection on the new surface 
obtained by blowing up a point. This procedure is particularly useful when the point in question is smooth. 
\item Given a surface which is a direct product of two curves, taking the exterior tensor product of the collections on the curves, and ordering it appropriately, gives a collection on the product space.
\end{itemize}  

Perhaps the most important tool in discovering an exceptional collection is
being able to check that it is indeed exceptional.  Taking advantage of
some recent mathematical results 
in local cohomology \cite{ToricCoh}, we have implemented a routine
in Macaulay2 which checks whether or not a given collection is exceptional.

The Macaulay2 routine, as any computer algorithm, only works on one example at a time. But one would like to treat an entire class of collections all at once.  Fortunately, after studying several examples by Macaulay2, we get a glimpse of the general structure in several cases. Then, using algebraic-geometric tools, we are able to prove that our collections on
$Y^{p,p-1}$, $Y^{p,p-2}$, and more generally $Y^{p,q}$, where $p-q$ is even,\footnote{There is another technical condition on $p$ and $q$, as discussed 
in Section~\ref{s:proofs}.} are exceptional for all $p$ and $q$.

To convert the collection into a quiver gauge theory, once we know it is exceptional,
we need to calculate all the maps between the sheaves.  Usually this calculation
is impossible to do by hand.  However, in our Macaulay2 routine the maps are a byproduct.  We have an
even faster Maple routine, which computes the same maps, but cannot always deal with the $\Ext^1$'s. 

The two routines are based on very different mathematics. The Maple routine 
counts lattice points inside polytopes, which is a linear programming problem. The Macaulay2 routine on the other hand uses a purely algebraic {\em local cohomology} calculation. Since this approach is completely new in the physics literature, we include a more detailed description of it. The fact that the two procedures {\em always} gave us the same answer for the $\Ext^0$'s and $\Ext^2$'s provides us with a strong consistency check.


In the next section, we introduce the mathematical machinery needed to write down collections of sheaves on these toric surfaces and check their exceptionality.  Then in sections 3, 4 and 5 we derive exceptional collections for a dozen or so specific
examples and from these collections calculate the gauge theory.  As a warm-up exercise, we rederive
collections for the toric del Pezzos, but now using toric techniques.  We also derive
collections for some simple ${\mathbb C}^3 / {\mathbb Z}_n$ orbifolds, finding
familiar quivers.  Finally, we look at the more complicated and interesting $L^{p,q,r}$
singularities of which the $Y^{p,q}$'s are a subset.  (We also consider the $X^{p,q}$ \cite{HananyXpq}.) 
Section 6 contains mathematical proofs that our collections for $Y^{p,p-1}$ and $Y^{p,p-2r}$ are exceptional.

In the last section of the paper, in an inversion of the philosophy espoused above,
we compute a quiver for the $L^{p,q,r}$ directly from the toric data.  We then
give an algorithm for converting the quiver into a collection we believe
should in general be strongly exceptional. 

\section{The Specifics}

\subsection{From Cones to Surface}

\label{s:conetosurface}

Our goal is to develop an algorithm for converting a toric Calabi-Yau cone into
a quiver gauge theory via exceptional collections. We take steps in this
direction, by providing methods for generating strongly exceptional collections
of line bundles from the toric diagram, and then a procedure for converting that collection into
a quiver gauge theory. 

Our starting point is the toric data for a local Calabi-Yau three-fold.  In other words,
we are given a {\em coplanar} collection of at least three vectors in ${\mathbb Z}^3$ .  For example, $V_1 = (1,1,0)$, $V_2 = (1,0,1)$ and $V_3=(1,-a,-b)$,  with $a$ and $b$ positive
integers,
describe an ${\mathcal N}=1$ supersymmetry preserving
 orbifold ${\mathbb C}^3/ {\mathbb Z}_{1+a+b}$.   These three vectors $V_i$ indeed end on the plane $x=1$. 

The next step is to convert
this three dimensional cone into a two dimensional fan describing a compact, possibly
singular toric surface.  
Assume all the vectors for the $3d$ cone
are written in the form $V_i = (1,*,*)$.
Truncating the first entry of the vector, we obtain a set of two dimensional vectors
$\{v_i\}$.  These $v_i$ define the vertices of a polygon.  
If this polygon contains one or more internal lattice points, then we can partially resolve
the $3d$ singularity by blowing up a four cycle.  

It is in this conversion of cone to surface that we are naturally led into the language of stacks.  The $v_i$ that arise in this procedure are often not primitive, i.e.~there is some positive integer
$n>1$ such that $v_i/n$ is also a lattice point.  These types of fans are simple,
two dimensional examples of the toric stacks described in \cite{BorisovDM}.  

The choice of the internal lattice point 
determines the origin of the two dimensional fan.  For example, for 
${\mathbb C}^3 / {\mathbb Z}_{1+a+b}$ and choosing the point $(1,0,0)$ as the origin
of the fan, we get the new set of vectors $v_1 = (1,0)$, $v_2 = (0,1)$, and $v_3 = (-a,-b)$
which define the weighted projective space ${\mathbb P}(1,a,b)$.
For large $a$ and $b$, there are generically many interior lattice points
and hence many choices of origin.  The gauge theory should be independent 
of our choice,
and in practice we choose the lattice point that gives the simplest surface.
(We will see this independence in some of the examples.)
Sometimes there is no lattice point
inside the polygon -- for example for the conifold -- in which case
this exceptional collection method needs modification.\footnote{In the case where
there is no surface to blow up, we expect to be able to blow up a curve
and write an exceptional collection for the curve.}

The fact that the fan descends
from a cone means that the fan must define the corners of a convex polygon.
For example, consider the Hirzebruch surfaces ${\mathbb F}_n$ which 
are described by the four vectors $v_1 = (1,0)$, $v_2 = (0,1)$,
$v_3 = (-1,n)$, and $v_4 = (0,-1)$.  For $n=0$ or $n=1$, these
four vectors define a convex quadrilateral from which
we can construct Calabi-Yau cones.  However for
$n>1$, the vector $v_2$ lies inside
the triangle defined by $v_1$, $v_3$ and $v_4$ and the Calabi-Yau cone
is better thought of as a ${\mathbb C}^3 / {\mathbb Z}_{2+n}$ orbifold.  

Starting with a Calabi-Yau cone, our problem can be rephrased
in terms of finding an exceptional collection for a special kind of compact toric surface
(or really toric stack).  
The surface $X$ is described by a fan of at least three vectors
in ${\mathbb Z}^2$, and in our case  the vectors describe a convex polygon.  
The vectors do not need to be primitive.

\subsection{Exceptional Collections}

An {\bf exceptional collection} of sheaves ${\mathcal E} = (E_1, E_2, \ldots, E_n)$ 
is an ordered set of sheaves which satisfy the following special properties:
\begin{enumerate}
\item Each $E_i$ is exceptional: $\Ext^q (E_i, E_i) = 0$ for $q>0$ and 
$\Ext^0(E_i, E_i) = \Hom(E_i, E_i) = {\mathbb C}$.
\item $\Ext^q (E_i, E_j) =0$ for $i>j$ and $\forall q$.
\end{enumerate}
In these notes, we will be most interested in the case where the collection is {\bf strongly
exceptional}, in which case $\Ext^q (E_i, E_j) = 0$ for $i<j$ and $q>0$.  
For smooth toric surfaces, the collection must be strong to generate a physical quiver
gauge theory \cite{Herzog1, AspinwallMelnikov}, 
and the same is true for singular surfaces as well.

For the most part, our sheaves can be thought of as line bundles, and 
line bundles are easy to describe in a toric context.\footnote{
For singular surfaces when $D$ is not a Cartier divisor, 
$\calo(D)$ is actually not a line bundle but only a reflexive sheaf. Nevertheless, for the sake of brevity  we don't emphasize this anymore.}  
For each ray $v_i$ in the fan,
there is a toric Weil divisor $D_i$.  The line bundles can then be expressed as
${\mathcal O} \left( \sum_i a_i D_i \right)$
for $a_i \in {\mathbb Z}$.
 One very special line bundle is the anti-canonical 
bundle:
\be
{\mathcal O}(-K) = {\mathcal O} \left (\sum_i D_i \right) \ .
\ee
As we said earlier, the Calabi-Yau cone is the total space of the canonical bundle over our surface.
The fact that our fan defines a convex polygon means that $K$ is negative.

To work with exceptional collections, we need to compute 
$\dim \Ext^q (E_i, E_j)$.  We are dealing with surfaces, and so the $q$'s of interest are $q=0$, 1, or 2.
For line bundles, the $\Ext$
groups can be understood in terms of sheaf valued cohomology, 
\be
\Ext^q_X ( {\mathcal O}(E), {\mathcal O}(F) ) =
H^q (X,   {\mathcal O}(F-E)) \ .
\ee
Serre duality\footnote{Every toric variety is Cohen-Macaulay \cite{Fulton}, and therefore the Grothendieck--Serre duality theory simplifies:  Grothendieck's dualizing complex is a just the canonical sheaf. } 
is useful in understanding what the non-vanishing $\Ext$ groups are
\be
H^q(X, {\mathcal O}(E) ) = H^{2-q} (X, {\mathcal O}( -E+ K))^\vee \  .
\ee

There is a straightforward way to compute $\dim \Hom(E_i, E_j)$ in the context of
toric geometry if $E_i$ and $E_j$ are line bundles \cite{Fulton}.  This dimension is the number of
global sections of the difference line bundle $E_j \otimes E_i^*$, which
we denote by $\O(D)={\mathcal O}(\sum_i a_i D_i)$.
To compute $\dim \Hom(E_i, E_j)$ one constructs the polygon
\be
\Delta_D=\{ u \in {\mathbb R}^2 : u \cdot v_i \geq - a_i \} \ .
\ee
for a two vector $u = (x,y)$.  The dimension $\dim \Hom(E_i, E_j)$ is the number of lattice points
in this polygon.  Although often tedious to compute by hand in all but the simplest cases,
it is easy to implement a point counting routine on a computer.  Through Serre
duality, this point counting routine allows one to compute $\dim \Ext^2(E_i, E_j)$ 
as well.

We also need to compute $\dim \Ext^1(E_i, E_j)$.  
There exist certain vanishing theorems which tell us when the higher cohomology vanishes.
For example, Kodaira's vanishing theorem says that given a line bundle 
$\calo(D)$ corresponding to an ample divisor $D$, then\footnote{
In the cases where $X$ has no worse than log terminal singularities (i.e. most of the cases discussed in this paper), 
there is a Kawamata-Viehweg vanishing theorem generalizing Kodaira vanishing.}
\be
\dim H^q(X, \calo(D \otimes K) )= 0\,,	\qquad \mbox{for any $q>0$.}
\ee

There is a straightforward way to test whether a toric divisor $D = \sum a_i D_i$ is ample.
Construct the piecewise linear function $\psi_D(v)$ 
where $v$ is a real valued vector in the space spanned
by the $v_i$ and the function $\psi_D$ takes the values $\psi_D(v_i) = -a_i$.
If $\psi_D$ is strictly convex, then $D$ is ample \cite{Fulton}.

In fact there is another useful result connected with this function $\psi_D$.  If
$\psi_D$ is convex, then the higher cohomology of $\calo(D)$ itself vanishes:
$H^q(X, \calo(D)) = 0$ for $q>0$.  However, since for us $K$ is negative,
Kodaira vanishing is stronger than this result.
 To keep things simple, 
in constructing a strong exceptional collection, we would like $E_i^* \otimes E_j$ to
be a line bundle corresponding to an ample divisor. 

We have implemented this lattice point counting routine and Kodaira vanishing
via $\psi_D$ on the computer algebra package Maple.  In the simple and especially
the smooth cases, this Maple routine is usually enough to tell whether the
collection is strongly exceptional.  However, in certain of the more complicated
singular examples, such as the $L^{p,q,r}$, Kodaira vanishing is not enough to
guarantee that $H^1=0$. In this case, we need to work harder.

In the appendix, we describe a more elaborate procedure based on local cohomology
for calculating all of $H^q$, including $H^1$, using Macaulay2.  This procedure is based on results
of \cite{ToricCoh}.

The exceptional collection remains exceptional (and generates the same quiver, since the quiver only
depends on differences)
if we tensor every object by the same line bundle. Physically this corresponds to large radius monodromy.
For example, ${\mathcal O}, {\mathcal O}(1), {\mathcal O}(2)$, and ${\mathcal O}(1), {\mathcal O}(2),
{\mathcal O}(3)$ are both exceptional collections on ${\mathbb P}^2$.  Therefore, let us choose
$E_1 = {\mathcal O}$.  Serre duality and Kodaira vanishing suggest a sense in which 
the collection is sandwiched between ${\mathcal O}$ and
${\mathcal O}(-K)$.

Given a strongly exceptional collection $\cale$, the  quiver gauge theory can be constructed from
the inverse collection $\cale^\vee$ (these are the fractional branes).  
The inverse collection is constructed by mutation, and is inverse
in the Euler character sense, $\chi(E_i, E_j^\vee) = \delta_{ij}$.  
The inverse collection is also exceptional although no longer strongly
exceptional.  The Euler character $\chi(E_i^\vee, E_j^\vee)$ can 
be interpreted as the number of arrows from node
$i$ to node $j$ in the quiver minus the number of arrows from node $j$ to node $i$
\cite{HerzogWalcher, AspinwallMelnikov}.

In the smooth case, a result of Kuleshov and Orlov \cite{KO} 
guarantees that there is at most
one map between any two objects in an exceptional collection and moreover
this map is either $\Ext^0$ or $\Ext^1$.  This result
appears no longer to be true in the singular case, and we have examples below.
As a result, computing $\chi$ only gives the quiver up to bidirectional arrows.
However, most of the time $\chi$ is enough.

Recently, Kawamata has given a construction for generating exceptional collections for any
toric variety  \cite{Kawamata}.  However, his exceptional collections are typically not strong and also
involve torsion sheaves.  Although we borrow his results for weighted projective space,
for more complicated examples, we follow our own simpler, but not so general, methods.

Presently, we review how to derive exceptional collections for some smooth toric surfaces, but first, we make a few remarks about ordering and the canonical class.

\subsection{Ordering the Sheaves}

\label{sec:ordering}

In the smooth (del Pezzo) case, the Kuleshov and Orlov result \cite{KO} discussed above means
it is enough to compute the Euler character to determine the maps between the
objects of an exceptional collection,
\be
\chi(E,F) = \sum_k (-1)^k \dim \Ext^k(E,F) \ .
\ee
For a smooth surface $\calb$, Riemann-Roch implies that
\begin{eqnarray}
\chi(E,F) &=& \int_{\calb} \ch(E^*) \ch(F) \Td(\calb) \\
&=& \rk(E) \rk(F) + \frac{1}{2}(\rk(E) \deg(F) - \rk(F) \deg(E) ) + \nonumber \\
&& \rk(E) \ch_2(F) + \rk(F) \ch_2(E) - c_1(E) \cdot c_1(F) \nonumber
\end{eqnarray}
where we have used the Chern character of the sheaf
\be
\ch(E) = (\rk(E), c_1(E), \ch_2(E)) \ .
\ee
Also, $\Td(\calb) = 1 - \frac{K}{2} + H^2$, where $K$ is the canonical class and $H$ 
is the hyperplane, with $\int_{\calb} H^2 = 1$.  Finally the degree $\deg(E) = (-K) \cdot c_1(E)$.

We denote $\chi_-(E,F)$ to be the antisymmetric part of $\chi$
\be
\chi_-(E,F) = \rk(E) \deg(F) - \rk(F) \deg(E) \ .
\ee
Note that for an exceptional pair $(E,F)$, $\chi_-(E,F) = \chi(E,F)$.  

From this last result, we can conclude that for a strongly exceptional collection, 
if $i > j$, then
\be
\frac{\deg(E_i)}{\rk(E_i)} \geq \frac{\deg(E_j)}{\rk(E_j)} \ .
\ee
Otherwise, we would get $\Ext^1$ maps.  This inequality motivates the definition of slope,
$\mu(E_i) = \deg(E_i)/\rk(E_i)$.  This way  the sheaves in a strongly exceptional collection
are ordered by their slope.  If we further insist that all the sheaves are line bundles 
($\rk(E) = 1$),
as we will do for most of this paper, the sheaves are ordered by their degree.

Although Riemann--Roch is substantially more complicated for our singular examples,
we have found on an example by example basis, given a strongly exceptional collection of line bundles, the bundles are strongly exceptional ordered by their degree.

We would like to introduce the notion of a helix before moving on to the examples in order
to try to bring a different point of view to the sandwiching of the collection mentioned above.
In the smooth case, Bondal and others \cite{Rudakov} have shown that given an exceptional
collection
\be
\cale = (E_1, E_2, \ldots, E_n) \ ,
\ee
we can generate another exceptional collection by right mutating $E_1$ all the way to the
end, $E_1 \to R_{E_n} \cdots R_{E_3} R_{E_2} E_1$.  Moreover, this right mutated
sheaf is
\be
R_{E_n} \cdots R_{E_3} R_{E_2} E_1 = E_1 \otimes (-K) \ .
\ee
This process can be inverted and repeated, generating a bi-infinite sequence of
sheaves called a helix, any $n$ adjacent elements of which form an exceptional collection.
We call any such $n$ adjacent elements a foundation for the helix.
In \cite{Herzog1}, it was shown that given a helix, any choice of foundation 
generates the same quiver.

Above we asserted that we needed a strong exceptional collection to generate
a physical gauge theory quiver.  In light of the existence of a helix, the requirement
is actually a little stronger.  The strong exceptional collection must generate a helix such
that any foundation is also strongly exceptional \cite{Herzog2}.  Such a helix is said to be strong.

From our Euler character formula, it is straightforward to see that a strong exceptional collection
in the smooth case generates a strong helix if and only if
\be
\mu(E_n) \geq \mu(E_1) + K^2 \ .
\ee
In the singular case, we have examples where this inequality does not hold but
we still get physical quivers.  Nonetheless, the inequality remains a good guideline, in the sense
that if the inequality is not satisfied and the collection is not strongly exceptional, then the collection 
is improved if some sheaves can be moved inside this window.  Given a strongly exceptional collection which
does not satisfy this inequality, in the singular cases we have studied, there is always
a related strongly exceptional collection which does and which produces the same quiver.

\section{Smooth Toric Surfaces}

Exceptional collections for ${\mathbb P}^2$, ${\mathbb P}^1\! \times {\mathbb P}^1$ and
${\mathbb P}^2$ blown up at one, two or three points can all be constructed by following a remarkably
simple rule.  Given a fan $\Delta = \{v_i \in {\mathbb Z}^2 : i=1,\ldots, n\}$ such that the 
$v_i$ are presented in clockwise or 
counterclockwise order, there exists an exceptional collection of the form\footnote{
This collection was described in \cite{Hille}.}
\be
\calo, \calo(D_1), \calo(D_1 + D_2), \ldots, \calo(\sum_{i=1}^{n-1} D_i) \ .
\ee
Let's see how this rule works for each of the smooth Fano toric surfaces listed above.

\subsection{${\mathbb P}^2$}

We can write the fan in the form $v_1 = (1,0)$, $v_2=(0,1)$, and $v_3 = (-1,-1)$.  These
three vectors satisfy the relation $v_1 + v_2 + v_3 = 0$, and all three divisors are
linearly equivalent.  Thus, the collection $\calo$, $\calo(D_1)$, $\calo(D_1+D_2)$
can be written in the more familiar form $\calo$, $\calo(H)$, $\calo(2H)$ given above.

For this collection the $\Ext^q(E_i, E_j)$ groups vanish for any $q>0$..  The matrix
$S_{ij} = \dim \Hom (E_i, E_j)$ takes the form
\be
S = \left(
\begin{array}{ccc}
1 & 3 & 6 \\
0 & 1 & 3 \\
0 & 0 & 1
\end{array}
\right) \ .
\ee
From the inverse of this matrix, we can construct the gauge theory quiver
\be
S^{-1} = \left(
\begin{array}{ccc}
1 & -3 & 3 \\
0 & 1 & -3 \\
0 & 0 & 1
\end{array}
\right) \ .
\ee
The above-diagonal entries of $S_{ji}^{-1}$ 
can be reinterpreted as the number of arrows from node $i$ to node
$j$ in the quiver.  A negative sign means the orientation of the arrow should be reversed.

We will discuss ${\mathbb P}^2$ blown up at one point and ${\mathbb P}^1 \times {\mathbb P}^1$ later, and 
in a slightly different way, which will be useful for our singular examples.  So, to avoid repetition,
we jump directly to $dP_2$.

\subsection{${\mathbb P^2}$ blown up at two points: $dP_2$}

A fan for $dP_2$ is given by 
\be
v_1 = (1,0), \, v_2 = (0,1),\, v_3 = (-1,0),\, v_4 = (-1,-1),\, v_5 = (0,-1) \ .
\ee
Using our rule, there is an exceptional collection of the form
\be
(0,0,0,0,0), (1,0,0,0,0), (1,1,0,0,0), (1,1,1,0,0), (1,1,1,1,0) \ ,
\ee
where we have introduced the shorthand notation 
\begin{equation}
\calo(\sum_i a_i D_i) \equiv (a_1, a_2, \ldots, a_n).
\end{equation}
The higher $\Ext$ groups vanish for this example and we find that
\be
S = \left(
\begin{array}{ccccc}
1 & 2 & 4 & 5 & 6 \\
0 & 1 & 2 & 3 & 4 \\
0 & 0 & 1 & 1 & 2 \\
0 & 0 & 0 & 1 & 1 \\
0 & 0 & 0 & 0 & 1
\end{array}
\right)
\ee
from which we calculate 
\be
S^{-1} = \left(
\begin{array}{ccccc}
1 & -2 & 0 & 1 & 1 \\
0 & 1 & -2 & -1 & 1 \\
0 & 0 & 1 & -1 & -1 \\
0 & 0 & 0 & 1 & -1 \\
0 & 0 & 0 & 0 & 1
\end{array}
\right) \ .
\ee
The quiver corresponding to $S^{-1}$ is precisely case II from page 13 of \cite{Hananyold}.

\subsection{$dP_3$ }

The fan in this case is
\be
(1,0), (1,1), (0,1), (-1,0), (-1,-1), (0,-1) 
\ee
from which we construct the collection
\be
(0,0,0,0,0,0), (1,0,0,0,0,0), (1,1,0,0,0,0), \ldots, (1,1,1,1,1,0) \ .
\ee
From this collection, we find that the higher $\Ext$ groups vanish
while
\be
S = \left(
\begin{array}{cccccc}
1 & 1 & 2 & 3 & 4 & 5 \\
0 & 1 & 1 & 2 & 3 & 4 \\
0 & 0 & 1 & 1 & 2 & 3 \\
0 & 0 & 0 & 1 & 1 & 2 \\
0 & 0 & 0 & 0 & 1 & 1\\
0 & 0 & 0 & 0 & 0 & 1
\end{array}
\right) \ .
\ee
The inverse matrix is then
\be
S^{-1} = \left(
\begin{array}{cccccc}
1 & -1 & -1 & 0 & 1 & 1 \\
0 & 1 & -1 & -1 & 0 & 1 \\
0 & 0 & 1 & -1 & -1 & 0 \\
0 & 0 & 0 & 1 & -1 & -1 \\
0 & 0 & 0 & 0 & 1 & -1\\
0 & 0 & 0 & 0 & 0 & 1
\end{array}
\right) \ .
\ee
The corresponding quiver was first written down using toric methods, and is Model I of Beasley and Plesser \cite{BeasleyPlesser}.

\section{Weighted Projective Spaces}	\label{s:wprsp}

Let's consider the weighted projective space ${\mathbb P}(1,a,b)$ defined by the complete
fan $v_1 = (-a,-b)$, $v_2 = (1,0)$ and $v_3 = (0,1)$.  These three vectors satisfy
the relation $v_1 + a v_2 + b v_3 = 0$, and thus the toric surface can be thought of as
a ${\mathbb C}^*$ quotient of ${\mathbb C}^3-\{0\}$, where the ${\mathbb C}^*$ acts 
with weights $(1, a, b)$ on the three coordinates.  

The total space of the canonical bundle ${\mathcal O}(K)$ over ${\mathbb P}(1,a,b)$
is a ${\mathbb Z}_{1+a+b}$ 
 orbifold of ${\mathbb C}^3$ with weights $(1,a,b)$.  String theorists 
(see for example \cite{Narayan}) 
associate a quiver
to such a ${\mathbb Z}_{a+b+1}$
 orbifold singularity.  The quiver consists of $a+b+1$ nodes and a collection of 
arrows between the nodes.  In particular, node $i$ will have three outgoing arrows, one
$X_i$ ending on node $i+1$, one $Y_i$ ending on node $i+a$ and one $Z_i$ ending on node $i+b$.
The quiver also has a number of relations.  In particular, consider two different paths that
begin and end on the same nodes.  These paths are equivalent if the two paths have the same number of $X$'s, the same number of $Y$'s, and the same
number of $Z$'s.  

We intend to rederive this quiver using an exceptional collection.  Kawamata \cite{Kawamata} shows
that a strong exceptional collection on ${\mathbb P}(1,a,b)$ will consist of all possible 
objects of the form
\be
{\mathcal O} \left(k_1 D_1 + k_2 D_2 + k_3 D_3 \right) \,,
\ee
such that $k_i  \geq 0$ and $0 \leq k_1 + a k_2 + b k_3 < 1+a+b$.  Moreover, given two objects
$E$ and $E'$ in the collection such that $E$ comes before $E'$, we must have
$k_1 + a k_2 + b k_3 < k_1' + a k_2' + b k_3'$.  
Thus, an exceptional collection on ${\mathbb P}(1,a,b)$ can always be written as
\be
{\mathcal O}, {\mathcal O}(D_1), {\mathcal O}(2D_1), \ldots, {\mathcal O}( (a+b)D_1) \ .
\ee
Note that the divisors satisfy the linear equivalence relations $aD_1 \sim D_2$ and $bD_1 \sim D_3$.
The same strongly exceptional collection is obtained by the authors of \cite{Auraux}.

Let's see how this claim
works in some specific examples.

\subsection{${\mathbb P}(1,1,2)$} 

This ${\mathbb Z}_4$ orbifold has an exceptional collection
\be
(0,0,0), (1,0,0), (1,1,0), (2,1,0)
\ee
where again  we use the shorthand notation ${\mathcal O}(k_1 D_1 + k_2 D_2 + k_3 D_3) \equiv 
(k_1, k_2, k_3)$.  From this collection the matrix
$S_{ij} = \dim \Hom (E_i, E_j)$ is
\be
S = \left(
\begin{array}{cccc}
1 & 2 & 4 & 6 \\
0 & 1 & 2 & 4 \\
0 & 0 & 1 & 2 \\
0 & 0 & 0 & 1 \\
\end{array}
\right) \ .
\ee

 From the gauge theory quiver for this ${\mathbb Z}_4$ orbifold, we can construct
the so-called Beilinson quiver on the toric surface ${\mathbb P}(1,1,2)$ by cutting
all the arrows that point from node $i$ to node $j$ where $i>j$ (see Fig.~\ref{p112}).  
These cut arrows are then reinterpreted as relations among the remaining maps. 
The dimensions in $S$ are the number of ways of getting from node $i$
to node $j$ taking into account the relations.

\FIGURE{
\epsfig{file=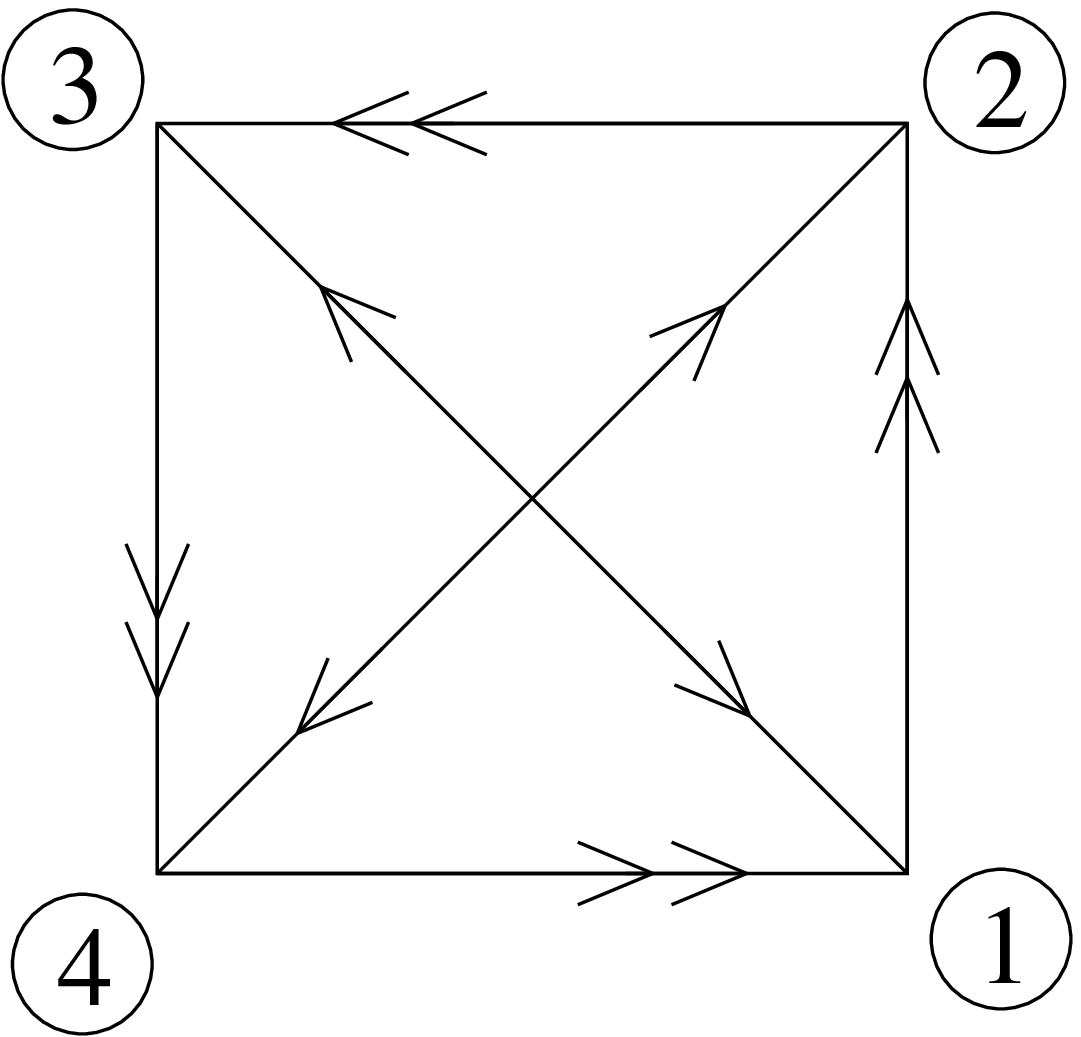, width=2in}
\epsfig{file=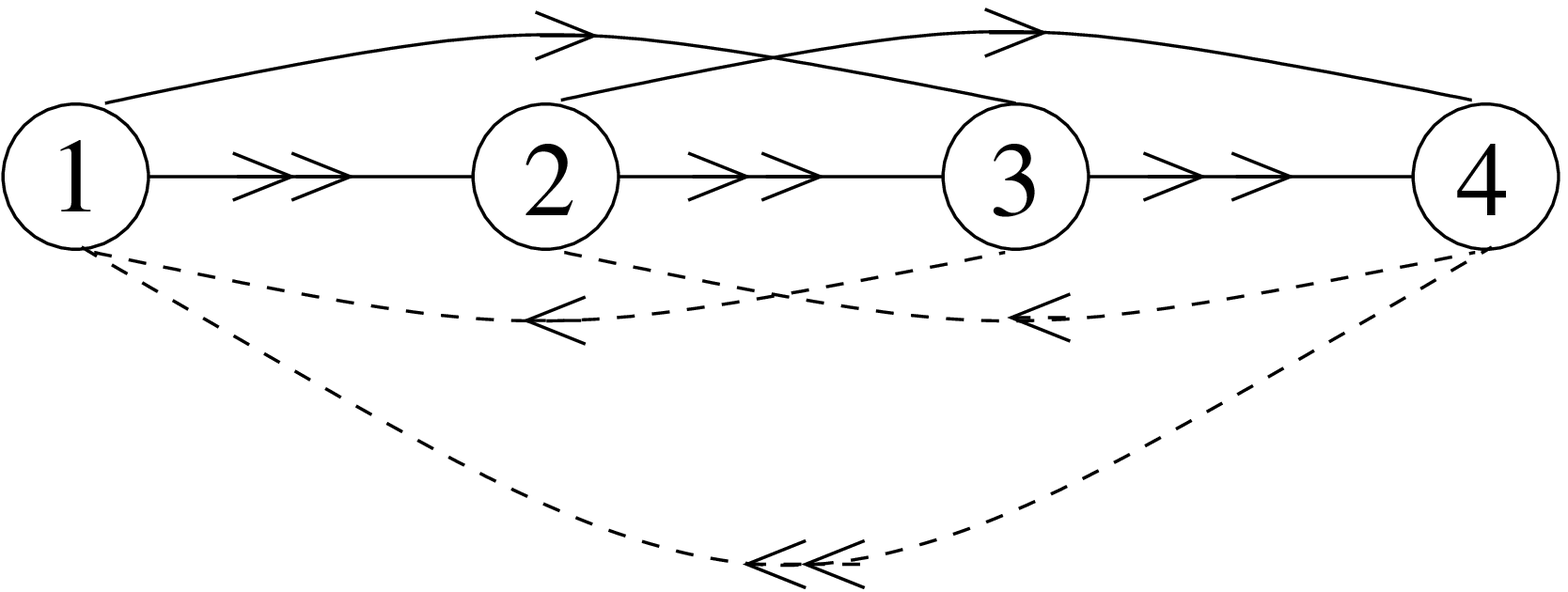, width=2.5in}
\caption{The gauge theory quiver and the Beilinson quiver for
${\mathbb P}(1,1,2)$.}
\label{p112}
}


Once again, the nonzero above the diagonal entries $S_{ij}^{-1}$ (for $i<j$) are  reinterpreted as the number of
arrows from $j$ to $i$ minus the number of arrows from  $i$ to $j$.  For
${\mathbb P}(1,1,2)$, we find
\be
S^{-1} = 
\left(
\begin{array}{cccc}
1 & -2 & 0 & 2 \\
0 & 1 & -2 & 0 \\
0 & 0 & 1 & -2\\
0 & 0 & 0 & 1 
\end{array}
\right)
\ee
which gives the quiver in Fig.~\ref{p112}.

Of course,  $S^{-1}$ doesn't show that there are in fact arrows running in opposite directions
between nodes 1 and 3 and also between nodes 2 and 4.  
In this computation $S_{ij}$ is the Euler character $\chi(E_i, E_j)$ because
the collection is strong.  $S_{ji}^{-1}$ is then the Euler character 
$\chi (E_i^\vee, E_j^\vee)$ of the dual gauge theory
collection \cite{HerzogWalcher, Herzog1, Herzog2}, which is not strong.

\subsection{${\mathbb P}(1,2,3)$}

This ${\mathbb Z}_6$ orbifold has an exceptional collection
 
\FIGURE{
\epsfig{file=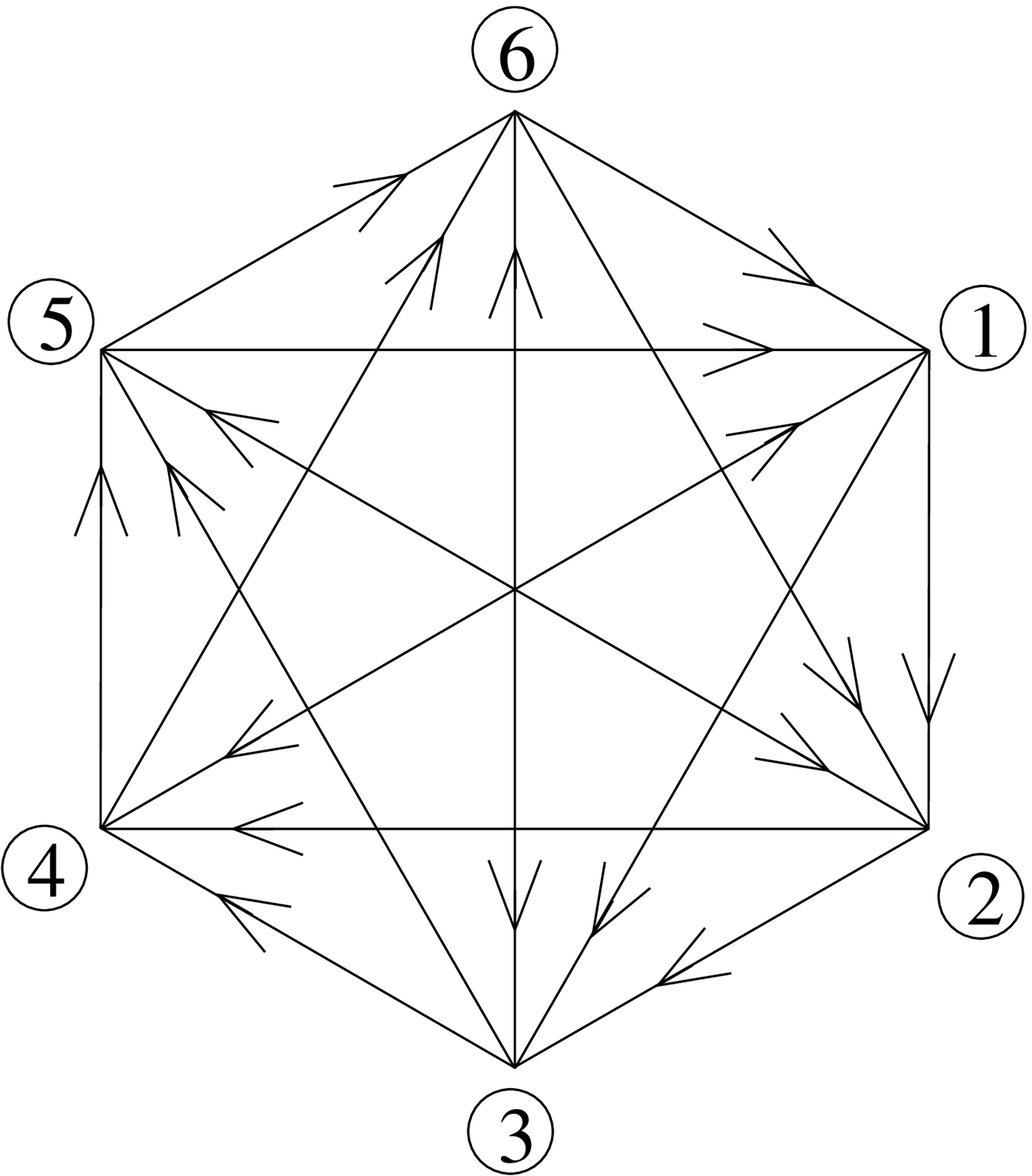, width=2.5in}
\caption{The ${\mathbb P}(1,2,3)$ gauge theory quiver.}
\label{p123}
}

\be
\begin{split}
&(0,0,0), (1,0,0), (2,0,0),\\
&(1,1,0),(1,0,1),(2,0,1) \, .
\end{split}
\ee

The $S_{ij}$ matrix here is
\be
S = 
\left(
\begin{array}{cccccc}
1 & 1 & 2 & 3 & 4 & 5 \\
0 & 1 & 1 & 2 & 3 & 4 \\
0 & 0 & 1 & 1 & 2 & 3 \\
0 & 0 & 0 & 1 & 1 & 2 \\
0 & 0 & 0 & 0 & 1 & 1 \\
0 & 0 & 0 & 0 & 0 & 1
\end{array}
\right) \ .
\ee
$S^{-1}$ then leads to the quiver in Fig.~\ref{p123}. Going the other way, it is again straightforward to reconstruct $S_{ij}$ from the Beilinson quiver.


\subsection{${\mathbb P}(1,a,a)$}		\label{s:p11a}

The weighted projective space $\P(1,a,a)$ can be thought of in two different ways. As a scheme or as a stack.\footnote{For a recent discussion see \cite{Auraux}.} 
Since the singularities at $(0:x_2:x_3)$ form a codimension one subspace, this space is not singular, even as a scheme, and hence ${\mathbb P}(1,a,a) \cong {\mathbb P}^2$. The origin of this ``smoothing-out'' is identical to the one that underlies the isomorphism ${\mathbb C} / {\mathbb Z}_n \cong {\mathbb C}$. Algebraically this isomorphism is
 in fact a trivial statement: $\Spec \C[x^n] \cong \Spec \C[x]$. We will return to this issue at length in Section~\ref{s:proofs}.

The traditional language of toric geometry \cite{Fulton} is not well suited to distinguish between ${\mathbb P}(1,a,a)$ and ${\mathbb P}^2$, and there are two different formalisms which can be introduced to make this distinction.  The first is that of boundary divisors.  We add the extrinsic information that there is a codimension one singularity along the divisor $D_1$,
which branches $a$ times: ${\mathbb P}(1,a,a)$ is equivalent to ${\mathbb P}^2$ with
boundary divisor $(a-1) D_1 /a$ \cite{Kawamata}.  The second, more powerful formalism, is the recently discovered theory of {\em toric Deligne-Mumford stacks} of Borisov,  Chen and Smith \cite{BorisovDM}.  
Regardless of the formalism, from the string theory point of view, this singular
surface is embedded in a Calabi-Yau space, the total space of its canonical sheaf, and the singularity structure is unambiguous. It is only when we try to analyze the $3d$ problem with $2d$ tools that we have to use more powerful techniques.

Although for the proofs in section \ref{s:proofs} we used stacky technology, for our Macaulay2 and Maple routines the fact that two of the weights have a gcd larger than one posed no extra complication.  The algorithms are the same when computing the dimensions of the cohomology of varieties with or without codimension one singularities, although the interpretations differ.

The collection on ${\mathbb P}(1,a,a)$, which will be useful in the next section, is
\be
(0,0,0), (1,0,0), (2,0,0), \ldots, (a-1,0,0), 
\ee
\[
(0,0,1), (1,0,1), (2,0,1), \ldots, (a-1,0,1) \ .
\]



\section{Beyond Weighted Projective Spaces}
\label{sec:tools}

As we alluded to in the introduction, at the moment we have only two tools in our basket for deriving exceptional
collections for toric surfaces that are generated by more than three rays.
The first is a divisorial blow-up and the second is an exterior tensor product.

For the divisorial blow-up, we start with two toric surfaces $X$ and $Y$ related by the blow-up map $\pi\!: X \to Y$.  Assuming that $${\mathcal E} = {\mathcal O}_Y, E_2, E_3, \ldots E_n$$ 
is an exceptional collection on $Y$, we would like to figure out the collection on $X$. 

Let the fan for $Y$ have $m$ rays.
The blow-up corresponds to adding 
a primitive ray $v_{m+1}$ to the fan, and also gives a new toric divisor $D_{m+1}$.  
The new
exceptional collection is taken to be
$${\mathcal E}' = {\mathcal O}_X, {\mathcal O}_X(D_{m+1}), \pi^* E_2, \ldots, \pi^* E_n .$$

But we can be much more explicit. We organize the rays in a counterclockwise fashion, and let's assume that the blow-up corresponds to the relation
$a_m v_m + a_1 v_1 = a_{m+1} v_{m+1}$ for some integers $a_i$.
From a theorem of Kawamata \cite{Kawamata},\footnote{We will implicitly prove this theorem in Section~\ref{s:xpq1}.} 
if $E_j = (k_1, k_2, \ldots, k_m)$, then
\be
\pi^* E_j = \left( k_1, k_2, \ldots, k_m, \lfloor \frac{a_m k_m + a_1 k_1}{a_{m+1}} \rfloor \right)\,.
\ee
For a smooth blow-up  $a_m = a_1 = a_{m+1} = 1$, and we will show that the procedure generates a strongly exceptional collection.

For the tensor product case we assume that the fan decomposes into ${\mathbb P}^1$ and something for
which we already know the exceptional collection.  For example, 
the fan consists of $v_1, v_2, \ldots v_m, (1,0), (-1,0)$ and we already know an 
exceptional collection ${\mathcal E}$ for the $v_1, v_2, \ldots v_m$ part.
Let 
\be
{\mathcal E} = (k^1_{1}, k^1_{2}, \ldots, k^1_{m}), (k^2_{1}, k^2_{2}, \ldots, k^2_{m}), \ldots,
(k^n_{1}, k^n_{2}, \ldots, k^n_{m})\ ,
\ee  
for some collection of integers $k^i_j$.
Then the collection on the whole space is
\begin{eqnarray*}
{\mathcal E}' &=& (k^1_{1}, k^1_{2}, \ldots, k^1_{m},0,0), (k^2_{1}, k^2_{2}, \ldots, k^2_{n},0,0), \ldots,
(k^n_{1}, k^n_{2}, \ldots, k^n_{m},0,0), \\
&&
\; \; (k^1_{1}, k^1_{2}, \ldots, k^1_{m},1,0), (k^2_{1}, k^2_{2}, \ldots, k^2_{m},1,0), \ldots,
(k^n_{1}, k^n_{2}, \ldots, k^n_{m},1,0)
\ .
\end{eqnarray*}

\subsection{Two Easy Examples: $\F_0$ and $dP_1$}		\label{s:dp1}

Here are two examples of smooth toric surfaces to which we can apply the techniques described
above, in a context where we already know the answers.

Take $\F_0={\mathbb P}^1\! \times {\mathbb P}^1$ for which the fan is 
$v_1 = (1,0)$, $v_2 = (0,1)$, $v_3 = (-1,0)$, and $v_4 = (0,-1)$.  An exceptional collection on 
a single ${\mathbb P}^1$, described by $v_1$ and $v_3$, is just $(0,0,0,0), (1,0,0,0)$.  Using
our procedure described above, a collection on the whole space is then
\be
(0,0,0,0), (1,0,0,0), (0,1,0,0), (1,1,0,0) \ .
\ee
Often this collection is written in a slightly different way.  Since $D_1 \sim D_3$ and
$D_2 \sim D_4$, we can suppress the last two entries and write instead
${\mathcal O}, {\mathcal O}(1,0), {\mathcal O}(0,1), {\mathcal O}(1,1)$, the well known
three-block exceptional collection on ${\mathbb P}^1\! \times {\mathbb P}^1$.

Take ${\mathbb P}^2$ blown up at a point, also known as the first del Pezzo surface,
$dP_1$, or the first Hirzebruch surface
$\F_1$.  The fan for ${\mathbb P}^2$ is $v_1 = (0,-1)$, $v_2 =  (1,1)$, $v_3 = (-1,0)$ 
to which we add the
primitive ray $v_4 = (-1,-1)$.  A well known collection on ${\mathbb P}^2$ is usually
written $\calo, \calo(1), \calo(2)$ which we translate into our toric notation as
\be
(0,0,0), (0,0,1), (0,1,1) \ .
\ee
Running our procedure, the collection on $dP_1$ should be
\be
(0,0,0,0), (0,0,0,1), (0,0,1,1), (0,1,1,1) \ .
\ee
Let $H$ be the hyperplane divisor of ${\mathbb P}^2$ and $E$ the exceptional
divisor on $dP_1$.  With a little work, one can show that $D_1 \sim D_3 \sim H-E$,
$D_4 \sim E$, and $D_2 \sim H$.  Then this collection can be written in the more familiar
notation
\be
\calo, \calo(E), \calo(H), \calo(2H) \ .
\ee

\subsection{Singular Examples}

In this and the following subsections, we will be interested in two large classes of examples.  The first
are the $Y^{p,q}$ spaces \cite{Benvenuti}, $0 \leq q \leq p$, which have the cone generated by the vectors
\be	\label{e:ypqpts}
V_1 = (1,0,0)\ , \; \; V_2 = (1, 1,0)\ ,\; \; V_3 = (1, p,p)\ , \; \; V_4 = (1,p-q-1,p-q)
\ee 
and the $X^{p,q+1}$ \cite{HananyXpq} which have the additional vector
\be
V_5 = (1, p-q-2,p-q-1) \ .
\ee

We start out by considering the simplest $Y^{p,q}$ space, namely $q=p$.  

\subsubsection{$Y^{p,p}$}	\label{sec:ypp}

In this case, we can drop the $V_1 = (1,0,0)$ vector because it lies on the line joining $V_2$ and $V_4$.  
We then convert this cone into a two dimensional fan by choosing one of the interior points to be the origin. We depicted a typical example in Fig.~\ref{f:fan44}, indicating all the interior lattice points.

\FIGURE{
\begin{xy} <1.2cm,0cm>:
 {\ar@{-} (-1,0);(1,0) }
,{\ar@{-}@{.>} (0,0);(4,4)    *+!LU{v_3=(4,4)}}
,{\ar@{-} (4,4);(-1,0)    *+!RU{v_4=(-1,0)}}
,{\ar@{-} (4,4);(1,0)    *+!LU{v_2=(1,0)}}
,{\ar@{-}@{.>} (0,-.5);(0,1.5) }
,{\ar@{-}@{.>} (-3,0);(3,0) }
,(0,0)*{\bullet}	,(1,1)*{\bullet}	,(2,2)*{\bullet}	,(3,3)*{\bullet}	,(4,4)*{\bullet}	,(1,0)*{\bullet}	,(-1,0)*{\bullet}
,(0,-.5)*+!U{v_1=(0,0)}
\end{xy}
  \caption{The toric fan of $Y^{4,4}$.}
  \label{f:fan44}
}

If we choose $(1,p-1,p-1)$ as the origin, the $2d$ fan is given by $v_2 = (-p+2, -p+1)$, $v_3 = (1,1)$, and $v_4= (-p,-p+1)$, which satisfies the single relation $v_2 + v_4 + (2p-2)v_3 = 0$.  In other words, we recover the 
weighted projective space ${\mathbb P}(1,2p-2,1)$ that we expect from \cite{Benvenuti, MartelliSparks}
and have already treated in Section~\ref{s:wprsp}.  The cone over $Y^{p,p}$ is well known to be the orbifold $\frac{1}{2p}(1,2p-2,1)=\frac{1}{2p}(1,-2,1)$.

But we could as well choose any of the interior points $(1,i,i)$, for any $1\leq i \leq p-1$, as the origin of the $2d$ fan (see Fig.~\ref{f:fan44}). For example, choosing $(1,1,1)$ we arrive at the projective space ${\mathbb P}(p-1,p-1,2)$.

Whether we choose the origin $(1,p-1,p-1)$ or $(1,1,1)$, the gauge theories had better be the same. For $p$ even, this is easy to see. With the origin $(1,p-1,p-1)$ we had the $\frac{1}{2p}(1,1,-2)$ orbifold of ${\mathbb C}^3$.  In the second case, with $(1,1,1)$ as the origin, we got the $\frac{1}{2p}(p-1,p-1,2)$ orbifold.  But these two actions are the same: by our assumption $p$ was even and hence $\gcd(2p,p-1)=1$, and thus the two spaces differ only by a choice of the $\Z_{2p}$ generator (if the first
generator is $\omega$, the generator of the second orbifold is $\omega^{p-1}$).

When $p$ is odd, the situation is more subtle.  With choice of origin $(1,1,1)$, the fan is
$v_2 = (0,-1)$, $v_3 = (p-1,p-1)$, and $v_4 = (-2,-1)$.  These $v_i$ do not span our
${\mathbb Z}^2$ lattice, covering only every other point.  Thus, the space is not
${\mathbb P}((p-1)/2, (p-1)/2, 1)$ but a ${\mathbb Z}_2$ orbifold of it.
Fortunately, Kawamata \cite{Kawamata} tells us how to handle this case as well.  We begin
by writing the usual exceptional collection for ${\mathbb P}((p-1)/2, (p-1)/2, 1)$:
\be
\calo, \calo(D_3), \calo(2D_3), \ldots, \calo((p-1)D_3) \ .
\ee
On this ${\mathbb Z}_2$ orbifold, there is also a torsion divisor, $(p-1)D_3/2 - D_4$, of order 2: 
$(p-1)D_3-2D_4$ is linearly equivalent
to zero.  Under the orbifolding, a line bundle $\calo(D)$ on ${\mathbb P}((p-1)/2, (p-1)/2, 1)$ pushes forward to a direct sum
$\calo(D) \oplus \calo(D+(p-1)D_3/2-D_4)$, 
and Kawamata proves (Theorem 3.5 of \cite{Kawamata}) that
the collection
\[
\calo, \calo((p-1)D_3/2-D_4), \calo(D_3), \calo((p+1)D_3/2-D_4), \ldots,
\]
\be
\calo((p-1)D_3), \calo(3(p-1)D_3/2-D_4)
\label{torsionEC}
\ee  
is strongly exceptional on the orbifold.

Let's consider the case $p=3$.  The choice of origin $(1,2,2)$ leads to the weighted projective
space ${\mathbb P}(1,1,4)$ and the expected quiver on this ${\mathbb C}^3 / {\mathbb Z}_6$
orbifold.  However, if we choose $(1,1,1)$ for origin, then we need to use
the exceptional collection (\ref{torsionEC}).  From our Maple routine, it is easy to compute
\be
S^{-1} =
\left(
\begin{array}{rrrrrr}
1 & 0 & -1 & -2 & 1 & 2 \\
0 & 1 & -2 & -1 & 2 & 1 \\
0 & 0 & 1 & 0 & -1 & -2 \\
0 & 0 & 0 & 1 & -2 & -1 \\
0 & 0 & 0 & 0 & 1 & 0 \\
0 & 0 & 0 & 0 & 0 & 1 
\end{array}
\right) \ ,
\ee
which, up to reordering of the nodes, yields the same quiver as the ${\mathbb Z}_6$ orbifold with weights $(1,1,4)$.

The choice of origin has a nice geometric interpretation. Adding the extra interior point blows up a divisor in the \CY , partially desingularizing it. Going from one interior point to another is a birational transformation composed of a blow-down followed by a blow-up. Since this is a K\"ahler deformation, we have reasons to expect that the derived categories are equivalent. This has already been proven for the conifold \cite{Brig:flop}, and more generally for quotient stacks \cite{Kawamata:DC}, but our case is even more general. 

In practice, we try to avoid blowing up points which lead to this extra subtlety of torsion divisors.  The reason is that our Macaulay2 program cannot handle torsion divisors directly. More generally, this subtlety occurs whenever the fan only spans a sublattice of the
${\mathbb Z}^2$ lattice on which it is written.  
 
\subsubsection{$Y^{p,p-1}$}	\label{s:pp-1}

Next we treat $q=p-1$.  Choosing the origin $(1,1,1)$, we find the fan
\be\label{e:con1}
v_1 = (-1,-1), \; \;  v_2 = (0,-1), \; \; v_3 = (p-1,p-1), \; \; v_4 = (-1,0)\,.
\ee
In this case, $v_2$, $v_3$ and $v_4$ form the weighted projective space ${\mathbb P}(p-1, 1, p-1)$, while $v_1$ is the exceptional divisor of a blow-up. Note that the point we blew up, corresponding to the large cone $\{v_2,v_4\}$, was smooth.
From our construction, we expect an exceptional collection for $Y^{p,p-1}$ to be
\begin{eqnarray}\label{e:coll1}
\cale &=& (0,0,0,0), (1,0,0,0), (0,0,1,0), (0,0,2,0), \ldots, (0,0,p-2,0),  \\
&& \; \; (1,0,0,1),(1,0,1,1), (1,0,2,1),
\ldots, (1,0,p-2,1), (2,1,0,1) \ . \nonumber
\end{eqnarray}

To familiarize ourselves with the $Y^{p,p-1}$ case we consider the first four: 
 $Y^{1,0}$, $Y^{2,1}$, $Y^{3,2}$ and $Y^{4,3}$.
The simplest non-trivial $Y^{p,p-1}$ space is $Y^{1,0}$. By (\ref{e:con1}) the \CY\ variety $Y^{1,0}$ is  ``the'' conifold, first studied in the AdS/CFT context in 
\cite{Klebanov-Witten,Morrison-Plesser, Kehagias}. The $Y^{1,0}$ case is degenerate from the $Y^{p,p-1}$ point of view, since it has no interior lattice points, and its desingularization is a small resolution. 

The next in line, $Y^{2,1}$, is the cone over $dP_1$, which we already considered in Section~\ref{s:dp1}. So let us consider the next two examples of $Y^{p,p-1}$-type. 

\FIGURE{
\epsfig{file=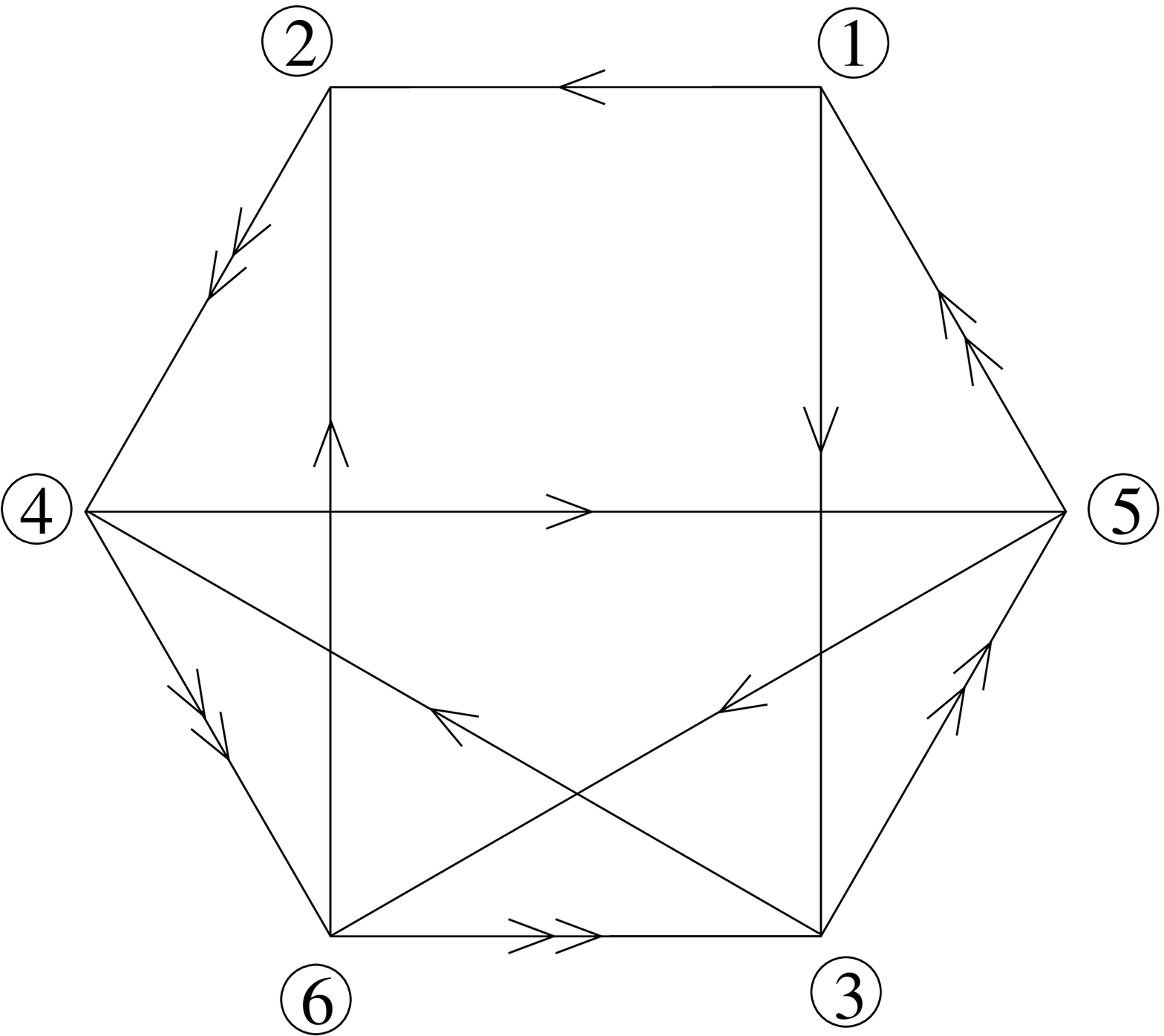, width=2.7in}
\caption{The usual quiver for $Y^{3,2}$.}
\label{Y32}
}

For $Y^{3,2}$, we find the collection
\be
\begin{split}
&(0,0,0,0), (1,0,0,0),(0,0,1,0),\\
&(1,0,0,1),(1,0,1,1),(2,1,0,1)  ,
\end{split}
\ee
from which we may compute
\be
S = 
\left(
\begin{array}{cccccc}
1 & 1 & 1 & 3 & 3 & 6 \\
0 & 1 & 0 & 2 & 2 & 5 \\
0 & 0 & 1 & 1 & 3 & 3 \\
0 & 0 & 0 & 1 & 1 & 3 \\
0 & 0 & 0 & 0 & 1 & 1 \\
0 & 0 & 0 & 0 & 0 & 1
\end{array}
\right)
\ee
and
\be
S^{-1} =
\left(
\begin{array}{rrrrrr}
1 & -1 & -1 & 0 & 2 & 0 \\
0 & 1 & 0 & -2 & 0 & 1 \\
 0 & 0 & 1 & -1 & -2 & 2 \\
 0 & 0 & 0 & 1 & -1 & -2 \\
 0 & 0 & 0 & 0 & 1 & -1 \\
0 & 0 & 0 & 0 & 0 & 1
\end{array}
\right) \ .
\ee

Using $S_{ij}^{-1}$, we may compute the usual quiver for $Y^{3,2}$ (see Fig.~\ref{Y32}).
Note this quiver is well split \cite{Herzog2} given the cyclic ordering from the exceptional 
collection.  
Well split means that if we order the nodes cyclically such that node $n+1$ is equivalent to
node 1, all the incoming arrows to node $i$ come from nodes $i-1$, $i-2$, etc. and all the 
outgoing arrows go to nodes $i+1$, $i+2$, etc.  Seiberg duality is so far understood
as a mutation only  for well split nodes.



Now onto $Y^{4,3}$: the collection (\ref{e:coll1}) gives the following matrices for $Y^{4,3}$
\be
S = 
\left(
\begin{array}{cccccccc}
1 & 1 & 1 & 1 & 3 & 3 & 3 & 6 \\
0 & 1 & 0 & 0 & 2 & 2 & 2 & 5\\
0 & 0 & 1 & 1 & 1 & 3 & 3 & 3 \\
0 & 0 & 0 & 1 & 1 & 1 & 3 & 3  \\
0 & 0 & 0 & 0 & 1 & 1 & 1 & 3 \\
0 & 0 & 0 & 0 & 0 & 1 & 1 & 1\\
0 & 0 & 0 & 0 & 0 & 0 & 1 & 1 \\
0 & 0 & 0 & 0 & 0 & 0 & 0 & 1
\end{array}
\right)
\ee
and
\be
S^{-1} = 
\left(
\begin{array}{rrrrrrrr}
1 & -1 & -1 & 0 & 0 & 2 & 0 & 0 \\
0 & 1 & 0 & 0 & -2 & 0 & 0 & 1\\
0 & 0 & 1 & -1 & 0 & -2 & 2 & 0 \\
0 & 0 & 0 & 1 & -1 & 0 & -2 & 2  \\
0 & 0 & 0 & 0 & 1 & -1 & 0 & -2 \\
0 & 0 & 0 & 0 & 0 & 1 & -1 & 0\\
0 & 0 & 0 & 0 & 0 & 0 & 1 & -1 \\
0 & 0 & 0 & 0 & 0 & 0 & 0 & 1
\end{array}
\right)
\ee
which yields the single impurity quiver for $Y^{4,3}$ (Figure 4 of \cite{Benvenuti}).

\subsubsection{$Y^{p,p-2}$}	\label{s:pp-2}

Next we consider the example $Y^{p,p-2}$.  Choosing again the origin of the fan 
$(1,1,1)$, we get
\be
\begin{split}
v_1 = (-1,-1), \; \;  v_2 = (0,-1), \\
v_3 = (p-1,p-1), \; \; v_4 = (0,1) \, .
\end{split}
\ee
For these surfaces, the total space is a direct product: $\P^1\!\times \P^1(1,p-1)$. The rays $v_2$ and $v_4$ ``form'' a $\P^1$, while $v_1$ and $v_3$ give the weighted projective space.  We can write an exceptional
collection for the orbifolded ${\mathbb P}^1$ using the weighted projective
space
techniques described above.  The exceptional collection on the whole space is then
\begin{eqnarray}\label{e:coll2}
\cale &=& (0,0,0,0), (0,0,1,0), (0,0,2,0), \ldots, (0,0,p-2,0), (1,0,0,0), \\
& & \; \; (0,0,0,1), (0,0,1,1), (0,0,2,1), \ldots,
(0,0,p-2,1), (1,0,0,1) \ . \nonumber  
\end{eqnarray}

The simplest example is $Y^{2,0}$, which in fact is the cone over $\F_0$, as discussed in Section~\ref{s:dp1}. The next simplest example is $Y^{3,1}$, for which we find the 
exceptional collection 
\be
(0,0,0,0),(0,0,1,0),(1,0,0,0),(0,0,0,1),(0,0,1,1),(1,0,0,1)\, .
\label{Y31double}
\ee

From the collection, we compute
\be
S = 
\left(
\begin{array}{cccccc}
1 & 1 & 2 & 2 & 2 & 4 \\
0 & 1 & 1 & 0 & 2 & 2 \\
0 & 0 & 1 & 0 & 0 & 2 \\
0 & 0 & 0 & 1 & 1 & 2 \\
0 & 0 & 0 & 0 & 1 & 1 \\
0 & 0 & 0 & 0 & 0 & 1
\end{array}
\right)
\ee
and
\be
S^{-1} = 
\left(
\begin{array}{rrrrrr}
1 & -1 & -1 & -2 & 2 & 2 \\
0 & 1 & -1 & 0 & -2 & 2 \\
0 & 0 & 1 & 0 & 0 & -2 \\
0 & 0 & 0 & 1 & -1 & -1 \\
0 & 0 & 0 & 0 & 1 & -1 \\
0 & 0 & 0 &  0 & 0 & 1 
\end{array}
\right) \ .
\ee
This $S^{-1}$ gives rise to the double impurity quiver for $Y^{3,1}$
\cite{Benvenutitoric}  (see Fig.~\ref{Y31}).  
Seiberg dualizing on node three or four yields the quiver with
two single impurities.  Seiberg duality corresponds to mutating $E_3$
all the way to the right or $E_4$ all the way to the left.

\FIGURE{
\epsfig{file=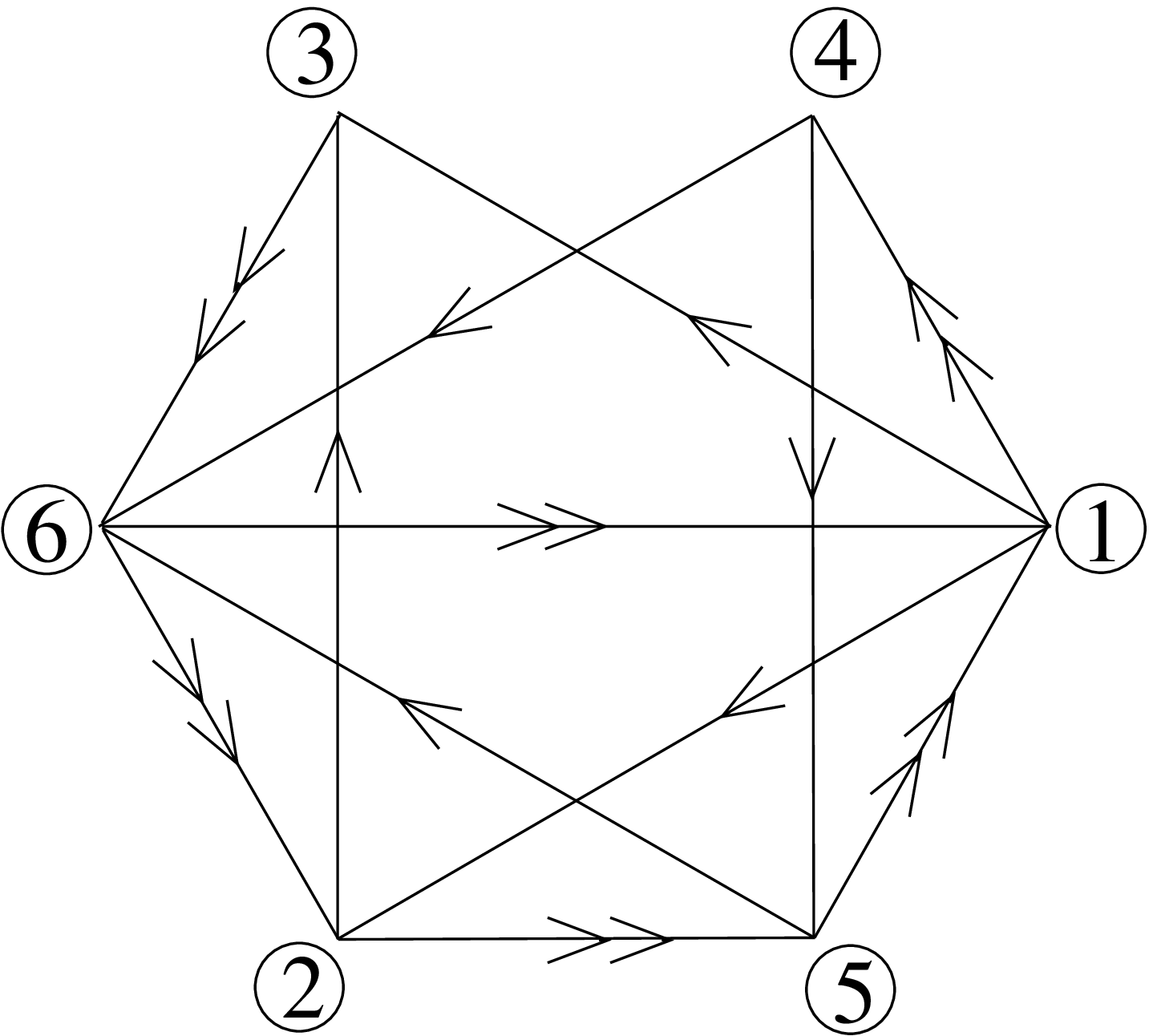, width=3in}
\caption{The double impurity quiver for $Y^{3,1}$.}
\label{Y31}
}


The next simplest case is $Y^{4,2}$. The  collection presented in (\ref{e:coll2}) gives the following matrix
\be
S^{-1} = 
\left(
\begin{array}{rrrrrrrr}
1 & -1 & 0 & -1 & -2 & 2 & 0 & 2 \\
0 & 1 & -1 & 0 & 0 & -2 & 2 & 0\\
0 & 0 & 1 & -1 & 0 & 0 & -2 & 2 \\
0 & 0 & 0 & 1 & 0 & 0 & 0 & -2  \\
0 & 0 & 0 & 0 & 1 & -1 & 0 & -1 \\
0 & 0 & 0 & 0 & 0 & 1 & -1 & 0\\
0 & 0 & 0 & 0 & 0 & 0 & 1 & -1 \\
0 & 0 & 0 & 0 & 0 & 0 & 0 & 1
\end{array}
\right)
\ee
which yields the standard double impurity quiver for $Y^{4,2}$ (Figure 2 of \cite{Benvenutitoric}).

We can also perform a Seiberg duality on this quiver to get two
single impurities, left mutating $E_5$ all the way to the left of the collection,
yielding
\be
E_5' = L_{E_1} E_5 = (0,0,0,-1).
\ee
Tensoring the whole collection by (0,0,0,1) then yields
\be
(0,0,0,0),(0,0,0,1),(0,0,1,1),(0,0,2,1),(1,0,0,1),(0,0,1,2),(0,0,2,2),(1,0,0,2) \ .
\label{Y42norm}
\ee
From this collection we find 
\be
S^{-1} = 
\left(
\begin{array}{rrrrrrrr}
1 & -2 & 0 & 0 & 0 & 1 & 0 & 1 \\
0 & 1 & -1 & 0 & -1 & 0 & 0 & 0\\
0 & 0 & 1 & -1 & 0 & -2 & 2 & 0 \\
0 & 0 & 0 & 1 & -1 & 0 & -2 & 2  \\
0 & 0 & 0 & 0 & 1 & 0 & 0 & -2 \\
0 & 0 & 0 & 0 & 0 & 1 & -1 & 0\\
0 & 0 & 0 & 0 & 0 & 0 & 1 & -1 \\
0 & 0 & 0 & 0 & 0 & 0 & 0 & 1
\end{array}
\right)
\ee
which gives a quiver with two single impurities and is well split (Figure 6 of \cite{Benvenuti} or
Figure 2 of \cite{Benvenutitoric}).

\subsection{Some Conjectures}

While we were able to motivate the collections for $Y^{p,p-1}$ and $Y^{p,p-2}$,
here we are doing something closer to guesswork, extrapolating results for small $p$.

For $Y^{p,p-3}$, and $Y^{p,p-4}$, the collection
\be\label{trig:mor}
\begin{split}
&(0,0,0,0), (0,0,0,1), (0,0,1,1), (0,0,2,1), \ldots (0,0,(p-2), 1), \\
&\qquad (0,0,1,2), (0,0,2,2), \ldots (0,0,(p-1),2), (0,0,(p-1), 3)
\end{split}
\ee
appears to generate appropriate quivers for all $p$.
In particular, for $Y^{p,p-4}$, the collection generates a quiver with
two single impurities separated by a double impurity.  For $Y^{p,p-3}$,
the quiver has three adjacent single impurities.  For $Y^{p,p-2}$, the
quiver has two single impurities separated by a normal unit.
 
 We can look at an explicit example, say $Y^{4,1}$. The fan is given by
\be
v_1 = (-1,-1), v_2 = (0,-1), v_3 = (3,3), v_4 = (1,2)
\ee
and the collection (\ref{trig:mor}) reads
\be
(0,0,0,0),(0,0,0,1),(0,0,1,1),(0,0,2,1),(0,0,1,2),(0,0,2,2),(1,0,0,1),(1,0,0,2) \ .
\label{Y41}
\ee
 From this collection, we find
\be
S^{-1} = 
\left(
\begin{array}{rrrrrrrr}
1 & -2 & 0 & 0 & 1 & 0 & 0 & 1 \\
0 & 1 & -1 & 0 & 0 & 0 & -1 & 0\\
0 & 0 & 1 & -1 & -2 & 2 & 0 & 0 \\
0 & 0 & 0 & 1 & 0 & -2 & 0 & 1  \\
0 & 0 & 0 & 0 & 1 & -1 & 0 & 0 \\
0 & 0 & 0 & 0 & 0 & 1 & -1 & 0\\
0 & 0 & 0 & 0 & 0 & 0 & 1 & -2 \\
0 & 0 & 0 & 0 & 0 & 0 & 0 & 1
\end{array}
\right)
\ee
which gives the standard quiver with three single impurities.

For $Y^{p,0}$ and $Y^{p,1}$, the collection
\be
(0,0,0,0), (0,0,0,1), (0,0,1,1), (0,0,1,2), \ldots (0,0,p-1, p-1), (0,0,p-1,p) 
\label{Yp0conj}
\ee
appears to be strongly exceptional and generates the appropriate quivers.

Later, we will give a conjectured exceptional collection for any toric
surface generated by four rays, including the $Y^{p,q}$ as special cases.  
With our rules thus far, the simplest $Y^{p,q}$ for which we have not given a collection is
$Y^{7,2}$, which has 14 gauge groups!

We also note that all of the $Y^{p,q}$ quivers appear to be well split, a property that does not hold
for the $L^{p,q,r}$ spaces we analyze below.

\subsection{$X^{p,q}$}	\label{s:xpq}

Given a collection for $Y^{p,q-1}$, it is a simple matter to generate a collection for $X^{p,q}$. Let $X_{p,q}$ be the surface underlying the 3-fold $X^{p,q}$, i.e., $X^{p,q}$ is the total space of the canonical sheaf on  $X_{p,q}$, and similarly let $Y_{p,q-1}$ underly the 3-fold $Y^{p,q-1}$. We can in fact arrange things in such a way that $X_{p,q}$ is a blow-up of $Y_{p,q-1}$. In particular, choosing (1,1,1) as the origin of the fan, one finds that $X_{p,q}$ is a blow-up of $Y_{p,q-1}$ at a smooth point. As described at the beginning of Section~\ref{sec:tools}, we know how to deal with this case. 

The $X_{p,q}$ fan is generated by the vectors
\be
\begin{split}
&v_1 = (-1,-1) \; , \; \; \; 
v_2 = (0,-1) \; , \; \; \;
v_3 = (p-1,p-1) \\
&v_4 = (p-q-1, p-q) \; , \; \; \;
v_5 = (p-q-2, p-q-1) \; . \; \; \;
\end{split}
\ee
$Y_{p,q-1}$ is described by the first four vectors, and $X_{p,q}$ is obtained by adding $v_5$. The smooth blow-up is reflected in the relation
\be
v_5 = v_4 + v_1 \ .
\ee
Given a strongly exceptional collection 
$${\mathcal E} = \O_{Y_{p,q-1}}, E_2, E_3, \ldots E_n$$ 
on $Y_{p,q-1}$ we will prove in Section~\ref{s:proofs} that collection
$${\mathcal E}' = \O_{X_{p,q}},  \O_{X_{p,q}}(D_5), \pi^*E_2, \pi^*E_3, \ldots, \pi^* E_n$$ 
is also strongly exceptional.

Let's see how all this works out for the example of $X^{42}$. The fan in this case is
\be
(-1,-1),(0,-1),(3,3),(1,2),(0,1) \ .
\ee
Starting with exceptional collection for $Y^{4,1}$ from (\ref{Y41}), we find 
the collection 
\be
(00000),(00001),(00011),(00111),(00211),(00122),(00222),(10012),(10023)
\ee
for $X^{4,2}$, which is strongly exceptional and yields the known quiver \cite{HananyXpq}.

\subsection{Other Examples: $L^{p,q,r}$}

We now turn to some more complicated examples, the $L^{p,q,r}$ spaces of \cite{HananyLpqr, ItalianLpqr, onemoreLpqr}.

\subsubsection{$L^{1,5,2}$}

Consider the cone
\be
V_1 = (1,1,3) \ , \; \; \; V_2 = (1,2,1) \ , \; \; \; V_3 = (1,0,0) \ , \; \; \; V_4 = (1,0,1)\, .
\ee
for which we choose the origin (1,1,2). The 2d fan is then generated by
\be
v_1 = (0,1) \ , \; \; \; v_2 = (1,-1) \ , \; \; \; v_3 = (-1,-2) \ , \; \; \; v_4 = (-1,-1) \, .
\ee
This fan satisfies the two relations
\be
v_1 + v_3 - v_4 = 0 \ , \; \; \; v_2 + v_3 + 3 v_1 = 0
\ee
We can thus think of the surface as ${\mathbb P}(1,1,3)$ with one
smooth point blown up, corresponding to the exceptional divisor $D_4$.  The collection we came up with is
\be
(0,0,0,0), (0,0,0,1),(0,0,1,1),(0,0,2,2),(1,0,0,1),(1,0,1,2)
\ee
which yields
\be
S^{-1} = 
\left(
\begin{array}{rrrrrr}
1 & -1 & -1 & 1 & 0 & 1 \\
0 & 1 & -1 & 0 & -1 & 1 \\
0 & 0 & 1 & -2 & 1 & -1 \\
0 & 0 & 0 & 1 & -2 & 1 \\
0 & 0 & 0 & 0 & 1 & -2 \\
0 & 0 & 0 & 0 & 0 & 1
\end{array}
\right)
\ee
in agreement with Fig.~7 of \cite{HananyLpqr}.

\subsubsection{$L^{1,7,3}$}

The cone for $L^{1,7,3}$ is
\be
V_1 = (1,1,0) \ , \; \; \; V_2 = (1,0,1) \ , \; \; \; V_3=(1,0,2) \ , \; \; \; V_4 = (1,3,3) \ .
\ee
From this cone, we choose the origin (1,1,1) and the 2d fan
\be
v_1 = (0,-1) \ , \; \; \; v_2=(-1,0) \ , \; \; \; v_3=(-1,1) \ , \; \; \; v_4=(2,2) \ .
\ee
This fan satisfies the two relations
$v_1 + v_3 - v_2 = 0$ and $2 v_3 + v_4 + 4 v_1 = 0$ and 
thus we think of the surface as ${\mathbb P}(4,2,1)$ blown
up at one smooth point.  Using our procedure, we write down
the collection
\be
(0,0,0,0),(0,1,0,0),(0,0,0,1),(0,0,0,2), (0,0,0,3), (0,0,0,4),(0,0,0,5),(0,0,0,6) \ .
\ee
from which we find
\be S^{-1} = 
\left(
\begin{array}{rrrrrrrr}
1 & -1 & -1 & 0 & 1 & 0 & 1 & 0 \\
0 & 1 & 0 & -1 & 0 & -1 & 0 & 1\\
0 & 0 & 1 & -1 & -1 & 1 & -1 & 1 \\
0 & 0 & 0 & 1 & -1 & -1 & 1 & -1  \\
0 & 0 & 0 & 0 & 1 & -1 & -1 & 1 \\
0 & 0 & 0 & 0 & 0 & 1 & -1 & -1\\
0 & 0 & 0 & 0 & 0 & 0 & 1 & -1 \\
0 & 0 & 0 & 0 & 0 & 0 & 0 & 1
\end{array}
\right)
\ee
in agreement with Fig.~20 of \cite{HananyLpqr}. We will consider more general $L^{p,q,r}$ spaces in Section~\ref{s:exconj}.




\section{Proofs}	\label{s:proofs}

In this section we provide proofs of several of our statements about strongly exceptional collections, for infinitely many families at once. Since the local \CY\ spaces in question are total spaces of  canonical sheaves on algebraic surfaces, one wonders to what extent is the singularity structure of the \CY\ variety captured by the geometry of the surface. In fact, one must retain more information than the intrinsic data of the two dimensional toric variety represented by the toric fan. We choose to retain this extra embedding information by using the language of stacks. We will review some elementary aspects of {\em quotient stacks} and {\em toric stacks} shortly. 

We could avoid stacks altogether, but they give a very convenient setting in which to do our calculations.
 We emphasize that stacks are useful because of our $2d$ perspective, but could be 
avoided if we kept a $3d$ description of the singularity. 

Without proofs or motivation, let us sketch a few of the, from our point of view, relevant aspects of quotient stacks. For a nice review we refer to \cite{Eric:DC2}. Let $X$ be an algebraic variety, and $G$ be a group acting on it. Giving a coherent sheaf $\ms F$ on the quotient stack $[X/G]$ is equivalent to giving  a coherent sheaf $\c F$ on X, together with a $G$-equivariant structure. The $G$-equivariant sheaves do not in general descend to the quotient space $X/G$. The sheaves on $X/G$ are precisely those $G$-equivariant sheaves, $\c F$, for which the stabilizer $G_x$ of any point $x\in   X$ acts trivially on the stalk ${\c F}_x$. 
In this sense, stacks allow for more general sheaves on the ``quotient space'' than varieties; we exploit this excess to write down our exceptional collections. 

For D-branes and quivers, our bread and butter is computing $\Ext$ groups. Moving into the stacky context, one has a very simple result. Let $\ms E$ and $ \ms F$ be two coherent sheaves on the stack $[X/G]$. If we represent them using two $G$-equivariant sheaves $\c E$ and $\c F$, then the stacky $\Ext$ is given by the invariant part of the ``old-fashioned'' $\Ext$:
\begin{equation}\label{e:stcoh}
\Ext^n_{[X/G]} ({\ms E, \ms F}) \,=\, \Ext^n_X (\c E, \c F) ^G\,.
\end{equation}

In recent years the stringy orbifold cohomology has been understood mathematically \cite{Chen:Ruan}. An algebraic version of this theory has been defined in the context of stacks \cite{Abramov:Vistoli}. More recently the orbifold Chow ring of a general toric Deligne--Mumford stack was obtained by Borisov, Chen and Smith (BCS) \cite{BorisovDM}. The BCS construction is completely combinatorial. In particular, Borisov, Chen and Smith give a procedure for associating a toric stack to a finitely generated abelian group $N$ (rather than a lattice), a fan $\Sigma$ in the free part of the  group (i.e., in the vector space $N\otimes_\Z \Q$), and a collection of group elements (these are points on the rays of the fan with decorations that are valued in the finite part of the abelian group). We will use the the BCS results in this section, and we refer for details of the construction and references to \cite{BorisovDM,Horj:Borisov}.

Let us look at the simplest non-trivial example of interest to us, which in fact will appear in the sequel: the weighted projective line $\P^1(a,b)$, for $a$ and $b$ {\em not} necessarily relatively prime integers. As a variety, or scheme, $\P^1(a,b)$ is isomorphic to $\P^1$, as we already discussed in Section~\ref{s:p11a}. We will consider the stack ${\mathbf P}^1(a,b)$ from two different points of view: first as a toric stack, and then as a quotient stack.

Torically it is somewhat inconvenient to talk about the variety $\P^1(a,b)$: once the lattice $\Z$ is given, the toric divisors are associated to the rays (the positive and negative axis), and not the non-minimal lattice points $-b$ and $a$ on the rays:  
\begin{equation}\nonumber
\begin{xy} <1.2cm,0cm>:
,{\ar (-1,0);(1,0)}	,{\ar (1,0);(-1,0)}
,(0,0)*{\bullet}	,(-1,-0.3)*{-1}	, 	,(1,-0.3)*{1}
,(-3,-0.3)*{-b}		,(-3,0)*{\comp}	,(2,-0.3)*{a}	,(2,0)*{\comp}	
,{\ar@{-}@{.>} (-5,0);(5,0) }
\end{xy}
\end{equation}
The  non-minimal lattice points, $-b$ and $a$, correspond to the weights. Once we turn to stacks, in the language of toric Deligne-Mumford stacks \cite{BorisovDM}, one can add extra decorations to each ray. In our case these are two elements $r_1$ and $r_2\in \Z_k$, for a given integer $k$. In Kawamata's language \cite{Kawamata} the decorations give rise to a boundary divisor $\frac{r_1-1}{r_1}D_1+\frac{r_2-1}{r_2}D_2$. In particular, if $d=\gcd(a,b) > 1$, then we can choose $k=d$, and think of $\P^1(a,b)$ as $\P^1(\frac{a}{d},\frac{b}{d})$ with $\Z_d$-valued decorations. In the BCS language this example is $N=\Z\oplus \Z_d$, the same fan as that of $\P^1$, and the choice of the two elements in $N$ being $(-b,r_1)$ and $(a,r_2)$.

Now let's see how ${\mathbf P}^1(a,b)$ is described as a quotient stack. In fact it is also true that $\P^1(a,b) \iso \P^1/\Z_a\! \times \Z_b$, where the group $\Z_a\! \times \Z_b$ acts on the homogenous coordinates $( x_0:  x_1)$ of $\P^1$ as $(\omega_a x_0: \omega_b x_1)$ --- the $\omega$'s are the appropriate roots of unity. But once again, the  $\Z_a\! \times \Z_b$ quotient is trivial on the variety. We can instead consider the stack $\c X = {\mathbf P}^1(a,b) \cong [\P^1/ \Z_a\! \times \!\Z_b]$. This replaces the local quotient singularities of the weighted projective space $\P^1(a,b)$ by stacky structures in the weighted projective stack ${\mathbf P}^1(a,b)$.

From the above discussion, the stack $\c X = [\P^1/ \Z_a\! \times \!\Z_b]$ has a richer category of coherent sheaves than the toric variety $\P^1/\Z_a\! \times \!\Z_b \iso \P^1$. To be more precise let $\pi\!: \c X \to X$ denote the canonical projection from the stack to the space. Let $\bar D_1$ and $\bar D_2$ denote the two toric divisors  of $X = \P^1(a,b)$, i.e., $x_1 =0$ resp. $x_2 =0$. They are of course linearly equivalent on $X = \P^1$. As explained in \cite{Kawamata}, the map $\pi\!: \c X \to X$ ramifies along the $D_i$'s, and there are divisors $\c D_1$ and $\c D_2$ on $\c X$ such that $\pi^* \bar D_i = s_i \c D_i$. The $ s_i$'s here correspond to the  $r_i$'s of the toric BCS language.

But there is another covering map, since we are quotienting  $\P^1$ by $\Z_a\! \times \!\Z_b$: $\P^1 \to {\mathbf P}^1(a,b)$ (see \cite{Kawamata:DC} for more details). We can summarize the two covering maps in the commutative diagram
\begin{equation}
\xymatrix{
\P^1  \ar[dr]_{\varpi}\ar[rr]^{} & &\P^1(a,b)\\
   & {\mathbf P}^1(a,b)  \ar[ur]_{\pi}& }
\end{equation}
If $D_1$ and $D_2$ denote the two toric divisors  of $\P^1$, i.e., $D_i=\operatorname{div}(x_i)$, then  $\varpi(D_i)=\c D_i$ are divisors on $\c X = {\mathbf P}^1(a,b)$. {\em But} $\varpi^* \c D_1=a D_1$ and $\varpi^* \c D_2=b D_2$. Nevertheless, there is an invertible sheaf $\O_{\c X}(\c H)$ on $\c X = {\mathbf P}^1(a,b)$ such that $\varpi^*\O_{\c X}(\c H)=\O_{\P^1}(1)$. This is the sense in which there are more sheaves on ${\mathbf P}^1(a,b)$ than on $\P^1$.

In the light of (\ref{e:stcoh}) the cohomology groups of $\O_{ \c X}(m\c H)$ are particularly easy to describe in the quotient stack language: the elements of $\H^0(\c X, \O_{ \c X}(m\c H))$ are degree $m$ polynomials in two variables $x_0$ and $x_1$, invariant under the $\Z_a\! \times \!\Z_b$ action. The group action is $( x_0:  x_1) \mapsto (\omega_a x_0: \omega_b x_1)$. In other words, these are the total degree $m$ polynomials in $x_0^a$ and $x_1^b$. This result is very familiar;  it is a trivial generalization of Cox's homogenous coordinate ring, which naively would not make sense for $a$ and $b$ having a common factor. In this context $x_0^a$ and $x_1^b$ play the role of the variables to which we would normally assign degrees $a$ resp. $b$. We can conveniently assemble this discussion into a single equation
\begin{equation}
\bigoplus_{m \in \N} \H^0(\c X, \O_{ \c X}(m\c H))\, = \, \C[u,v]\,,
\end{equation}
where $\C[u,v]$ is a $\Z$-graded polynomial ring with $\deg\, u =a$ and $\deg\, v =b$.

Although our example was almost trivial, there is much more to toric stacks than meets the eye. 
We need and use these stacks in our proofs.
We start out by proving that the conjectured collection for $Y^{p,p-1}$ is strongly exceptional for all $p$. Next we prove the same thing for any $Y^{p,q}$ space for which $p-q$ is even, say $p-q=2r$, and $\gcd(p,r)=1$. Finally, we extend these results to the corresponding $X^{p,q}$ spaces.

\subsection{$Y^{p,p-1}$}


Let us start with the collection of line bundles listed in Eq.~(\ref{e:coll1}). Using linear equivalence we can bring it to a form that is more suitable for proving that it is strongly exceptional for all $p$. The toric divisors of $Y^{p,p-1}$ satisfy the following relations:
\begin{equation}
D_4 \sim D_2,\quad D_1+D_2 \sim (p-1)D_3\,
\end{equation}
Using these linear equivalences, the collection (\ref{e:coll1}) becomes
\begin{equation}
\O_X,\, \O_X(D_1),\, \O_X(D_3),\, \O_X(2D_3),\, \ldots\,,\O_X((2p-2)D_3)\,.
\end{equation}

Now we should recall from Section~\ref{s:pp-1} the basic geometry of the $Y^{p,p-1}$ spaces. With our choice of the origin in the  $x_1=1$ plane, $Y^{p,p-1}$ is the total space of the canonical sheaf over a complex surface $X_{p,p-1}$, which is the blow-up of the smooth point on the weighted projective space $Y_{p,p-1} = {\mathbb P}(p-1, 1, p-1)$.\footnote{The reader should {\em not} confuse $Y^{p,p-1}$ with $Y_{p,p-1}$. The former is a 3-fold, while the latter is a surface.}
Let $\pi\! : X_{p,p-1}\to Y_{p,p-1}$ be the blow-up map, $x$ be the point that is blown up, and let  $E=D_1$ be the exceptional divisor of the blow-up. Since $x$ is a smooth point we have that $E^2=-1$, and furthermore, in this case, $E= \P^1$.

The reader is now fully aware of the subtlety in handling $Y_{p,p-1} = {\mathbb P}(p-1, 1, p-1)$, and the previous paragraph as it stands is incomplete. We will use the fact that $\O,\, \O(1),\,  \ldots\,,\O((2p-2))$ is a strongly exceptional collection on the stack ${\mathbf P}(p-1, 1, p-1)$. Of course this statement would fail for the variety ${\mathbb P}(p-1, 1, p-1)\iso \P^2$. If we denote the minimal toric divisors of ${\mathbb P}(p-1, 1, p-1)$ by $D_i$, then the natural map $\pi\!: {\mathbf P}(p-1, 1, p-1) \to {\mathbb P}(p-1, 1, p-1)$ ramifies along $D_2$. The blow-up on the other hand took place in the atlas corresponding to the cone formed by $D_1$ and $D_3$. This mismatch, or avoidance if you will, allows us to use both classical algebraic geometry and the stacky exceptional collection. As we already said, one could forget about stacks, and instead keep track of the  $\Z_{p-1}\! \times \Z_{p-1}$--equivariant structure, but we find it more convenient to work in this hybrid setting. In particular, the theorems of classical algebraic geometry still apply in this subsection.

Now we are in a position to state the main result of this subsection:

\begin{prop}		\label{prop61}
The collection of rank $1$ sheaves $\O_X,\, \O_X(D_1),\, \O_X(D_3),\, \O_X(2D_3),\, \ldots\,,$ $\O_X( (2p-2) D_3)$, on the complex surface $X=X_{p,p-1}$ is strongly exceptional.
\end{prop}
\begin{proof}
We prove the proposition by computing the $\Ext$ groups  between any two elements in the collection, including any  element with itself. We start with $\O_X$ and $\O_X(D_1)$,\footnote{We drop the subscript $(p,p-1)$ from $X_{p,p-1}$, and similarly for $Y_{p,p-1}$.} and use a direct method. For the rest of the collection we derive some general results, which will be of use later on as well.

To start, we consider the short exact sequence (SES)\footnote{We abbreviate ``short exact sequence'' as SES, and  ``long exact sequence'' as LES.}
\begin{equation}\label{e1}
\ses{\O_X(-E)}{\O_X}{\O_E }\,.
\end{equation}
The associated cohomology long exact sequence (LES) implies that  $\H^i(X, \O_X(-E) ) = 0$, for  $i\geq 0$. 

Tensoring by the locally free, hence flat, sheaf $\O_X(E)$ the SES (\ref{e1}) gives another SES
\begin{equation} \label{e2}
\ses{\O_X}{\O_X(E)}{\O_E(-1) }\,,
\end{equation}
where we used the fact that $E^2=-1$. Now $\O_\P^1(-1)$ is a remarkable sheaf in that $\H^i(\P^1, \O_\P^1(-1) ) = 0$, for all $i$. Therefore the associated cohomology LES implies that $\H^0(X, \O_X(E) ) = \H^0(X, \O_X ) = \C$ and $\H^i(X, \O_X(E) ) = \H^i(X, \O_X ) = 0$, for  $i\geq 1$. 

Thus, so far, we have established that $ \O_X$ and $ \O_X(E)$ form a strongly exceptional collection. To compute the $\Ext$'s between the $\O_X(mD_3)$'s we exploit the fact that they arise from an exceptional collection on $Y$, via the pull-back $\pi\! : X \to Y$. More precisely we have that 
\begin{equation}
\O_X(D_3) = \pi^*\O_Y(1)\,, \quad \mbox{and consequently}  \quad \O_X(mD_3) = \pi^*\O_Y(m)  \quad \mbox{for any $m\in\Z$ }\,.
\end{equation}

The first question is therefore to compute $\Ext^i_X(\pi^*\O_Y(a), \pi^*\O_Y(b))$, for $a,b\in \Z$. We will prove a Parseval-type identity
\begin{lemma}	\label{prop6lemma}
${\mytext{Ext}}^{\,i}_X(\pi^*\O_Y(a),\pi^*\O_Y(b)) = {\mytext{Ext}}^{\,i}_Y(\O_Y(a),\O_Y(b))$, for all $a,b\in \Z$.
\end{lemma}
\begin{proof}[Proof of the Lemma]
The local freeness of the $\pi^*\O_Y(a)$'s reduces the problem to computing $\H^i(X,\pi^*\c F)$, where $\c F$ is a line bundle. For this we can use the ubiquitous {\em Leray spectral sequence}, which for a map $f\!:X\to Y$, and a sheaf $\c E$ on $X$ reads:
\begin{equation}\label{e:lss1}
E_2^{p,q}=\H^p(Y,\,R^q f_*\,{\c  E})\Longrightarrow \H^{p+q}(X,\, {\c  E})\,.
\end{equation}
For us $\c E = \pi^*{\c  F}$, and the Leray spectral sequence becomes
\begin{equation}\label{e:lss}
E_2^{p,q}=\H^p(Y,\,R^q \pi_*\,\pi^*{\c  F})\Longrightarrow \H^{p+q}(X,\, \pi^*{\c  F})\,.
\end{equation}
But we can trivially rewrite $R^q \pi_*\,\pi^*{\c  F}$ as $R^q \pi_*(\pi^*{\c  F}\otimes \O_X)$. Now we can use the projection formula\footnote{For a morphism of schemes $f\!:X\to Y$, a coherent sheaf $\c F$ on $X$, and a locally free sheaf  $\c E$ on $Y$, the projection formula is the isomorphism $R^i f_*\,({\c  F}\otimes f^* {\c  E} ) \iso R^i f_*\,{\c  F}\otimes {\c  E} $. (See, e.g., III.9 of \cite{Hartshorne:}.) } 
since our $\c F$ is a line bundle and get ${\c  F}\otimes R^q \pi_*\O_X$. But $\pi$ is a smooth blow-up of a smooth surface, and Proposition 3.5 in Section V.3 of \cite{Hartshorne:} tells us that $R^q \pi_*\O_X = \O_Y$ for $q=0$ and $0$ otherwise. The same conclusion can be drawn using a general result about birational toric maps (see, e.g., page 76 of \cite{Fulton}). So the Leray spectral sequence (\ref{e:lss}) degenerates and $\H^{p}(X,\, \pi^*{\c  F}) \cong \H^{p}(Y,\, {\c  F}) $, as promised.
\end{proof}

Using the lemma, and the fact that the standard exceptional collection $\langle \O_Y,\O_Y(1),\ldots $, $\O_Y(2p-2) \rangle$ on $Y ={\mathbb P}^2(p-1, 1, p-1)$ is strong, we can immediately conclude that  $\langle \O_X, \O_X(D_3), \ldots\ ,\O_X((2p-2)D_3)\rangle$ is a strong exceptional collection on $X$.

Finally we turn our attention to gluing the two  strong exceptional collections on $X$: $\langle \O_X, \O_X(E)\rangle$ and $\langle \O_X, \O_X(D_3), \ldots\ ,\O_X((2p-2)D_3)\rangle$. Therefore we need to evaluate the $\Ext^i_X(\pi^*\O_Y(a),\O_X(E))$'s, which boils down to computing $\H^i(X, \O_X(\pm E+mH))$ where for brevity we denoted $D_3$ by $H$. This notation also reminds us of the fact that $D_3$ originates from the hyperplane divisor on $Y ={\mathbb P}^2(p-1, 1, p-1)$ and inherits certain properties. 

Again we use some convenient SES's. First, tensoring (\ref{e1}) by $ \O_X(mH)$ gives\footnote{Here we used again a projection-type formula, but now for the intersection of divisors on a smooth surface: $E\cdot H = E\cdot \pi^* L = \pi_*E\cdot  L = \mbox{point}\cdot  L =0$, where $L$  is the hyperplane divisor on $Y$, i.e., $\O_Y(1)=\O_Y(L)$.}
\begin{equation}\label{e11}
\ses{\O_X(mH-E)}{\O_X(mH)}{\O_E }\,.
\end{equation}
From the  associated cohomology LES $\dim \H^0(X, \O_X(mH-E))= \dim \H^0(X, \O_X(mH))-1$,\footnote{Negative dimension means trivial vector space.} 
while $ \H^i(X, \O_X(mH-E))= \H^i(X, \O_X(mH))$ for $i>0$. By the lemma above $\H^i(X, \O_X(mH))=\H^i(Y, \O_Y(mH))$. The strong exceptional property of the collection on $Y$ guarantees that the latter one is $0$ for any $m\geq 0$ and $i>0$.

Finally, tensoring (\ref{e2}) by $ \O_X(mH)$ gives
\begin{equation}\label{e12}
\ses{\O_X(mH)}{\O_X(mH+E)}{\O_E(-1) }\,.
\end{equation}
In cohomology this translates into $ \H^i(X, \O_X(mH+E))= \H^i(X, \O_X(mH))$ for all $i\geq 0$. By the lemma, and the strong exceptionality on $Y$, these groups are all $0$ for any $-(2p-1)\leq m < 0$. Some bookkeeping now shows that this in fact completes the proof.
\end{proof}

\subsection{$Y^{p,p-2}$ and  $Y^{p,p-2r}$, for $\gcd(p,r)=1$}


It is natural to ask the question of when do the four points 
\be
V_1 = (1,0,0)\ , \; \; V_2 = (1, 1,0)\ ,\; \; V_3 = (1, p,p)\ , \; \; V_4 = (1,p-q-1,p-q)
\ee 
from Eq.~(\ref{e:ypqpts}), defining the $Y^{p,q}$ spaces, give rise to a surface that is a direct product of two curves. From the ``topology'' of the pointset it is immediate that a necessary and sufficient condition is to have the line connecting  $V_1$ and $V_3$ intersect the line connecting $V_2$ and  $V_4$ in a {\em lattice point}. It is elementary to see that the two lines intersect at $(\frac{p-q}{2},\frac{p-q}{2})$, hence we require that 
\begin{equation}
p-q = 2r\,, \qquad \mbox{for some $r\in \N.$}
\end{equation}
Choosing $(\frac{p-q}{2},\frac{p-q}{2})=(r,r)$ as the origin of the $2d$ fan we have the four vectors $v_1=(-r,-r)$, $v_2=(1-r,-r)$, $v_3=(p-r,p-r)$ and $v_4=(r-1,r)$. We depicted these four points and connecting lines in Fig.~\ref{f:fan}. 

\FIGURE{
\begin{xy} <1.2cm,0cm>:
 {\ar (0,0);(-2,-2) *+!R{v_1=(-r,-r)}}
,{\ar (0,0);(3,3)    *+!LU{v_3=(p-r,p-r)}}
,{\ar (0,0);(-1,-2)  *+!LU{v_2=(1-r,-r)}}
,{\ar (0,0);(1,2)    *+!RD{v_4=(r-1,r)}}
,{\ar@{-}@{.>} (0,-1.5);(0,1.5) }
,{\ar@{-}@{.>} (-3,0);(3,0) }
,(-1.2,-1.7)*+[o][F.]{{\mathcal C}_1}
,(1.2,1.7)*+[o][F.]{{\mathcal C}_3}
,(1,-1)*+[o][F.]{{\mathcal C}_2}
,(-1,1)*+[o][F.]{{\mathcal C}_4}
\end{xy}
  \caption{The toric fan of $Y^{p,q}$, for $p-q=2r$ .}
  \label{f:fan}
}

From the toric fan in Fig.~\ref{f:fan} it is now intuitively clear that $v_1$ and $v_3$ form a weighted projective space $\P^1(p-r,r)$, while $v_2$ and  $v_4$ form the projective space $\P^1$.\footnote{Note that there is no  lattice point on the ray connecting the origin to $v_2=(1-r,-r)$ closer to the origin.}
This fact can be see even more explicitly if we use Cox's holomorphic quotient construction,\footnote{See Appendix~\ref{a:app1} for a brief review.} and once again we will have to deal with  some stacky subtleties.

Let us work this out in detail. The space $Y^{p,q}$ is the total space of the canonical sheaf over a complex surface $X_{p,q}$. Fig.~\ref{f:fan} gives the toric fan of $X_{p,q}$. The linear equivalence relations among the toric divisors of $X_{p,q}$ are
\begin{equation}\label{e:weights}
D_4 \sim D_2,\quad r D_1 \sim (p-r)D_3\,.
\end{equation}
Let $x_1,\ldots , x_4$ be coordinates on $\C^4$. Following Cox \cite{Cox:HoloQout}, $X_{p,q}$ is the quotient of $\C^4-\{x_1=x_3=0,x_2=x_4=0\}$ by $\C^*\times \C^*$, where the weights of the two $\C^*$ actions can be read out from (\ref{e:weights}): $(\lambda_1^{p-r} x_1, \lambda_2 x_2, \lambda_1^r x_3, \lambda_2 x_4)$, where $( \lambda_1 , \lambda_2)\in \C^*\times \C^*$. 

Consider for a moment the $x_1=0$ subspace, call it $\c D_1$. Since the subset $\{x_1=x_3=0\}$ is excluded and  $x_1=0$ we must have $x_3\neq 0$. Therefore we can use the first $\C^*$ action to completely fix $x_3$, say to $x_3=1$. What we are left with is $\C^2-\{x_2=x_4=0\}/\C^*\iso \P^1$. But we ignored the fact that $\C^*$ acts on $x_3$ with weight $r$. So we have to be more careful here, and see if there is any stacky subtlety. 

To determine the ``topology'' of $\c D_1$ as a stack we can use the toric  method of \cite{BorisovDM}. This is actually interesting in itself, since we start with a toric variety with $N=\Z^2$ and no torsion, but for the subspace $\c D_1$ the corresponding $N_{\c D_1}$ will have torsion. 

More generally, if we consider the abelian group $\Z^2$, and the subgroup $\langle (a,b)\rangle$ generated by $(a,b)$, for $a,b \in \Z$, then the quotient $\Z^2/\langle (a,b) \rangle$ is again a finitely generated abelian group. By the fundamental theorem of finitely generated abelian groups it is necessarily of the form $\Z^r\oplus \mbox{finite torsion}$. In fact it is easy to show that $\Z^2/\langle (a,b)\rangle\iso \Z\oplus \Z_{\gcd(a,b)}$. Using this it is automatic that $N_{\c D_1}\iso \Z\oplus \Z_r$. The $Star$ of $\c D_1$ consists of the cones $\c C_1$ and $\c C_4$. The image of $v_2=(1-r,-r)$ in $N_{\c D_1}\iso \Z\oplus \Z_r$ is $(1-r,-r)\equiv (1,0) \mod v_1$. Similarly, the image of $v_2=(r-1,r)$ in $N_{\c D_1}\iso \Z\oplus \Z_r$ is $(-1,0)$. Running the BCS construction one obtains a $\P^1$, with trivial $\Z_r$ action. 

The fact that $\c D_1=\P^1$ is only intuitively possible to be seen from Cox's construction, since the latter one gives a variety, rather than a stack. A similar analysis applies to the $x_3=0$ subspace, and we get another trivially-stacky $\P^1$. 

The $x_2=0$ and $x_4=0$ subspaces are  more straightforward. Consider the subspace $\c D_2$ given by $x_2=0$. Once again, the $\{x_2=x_4=0\}$ subset is excluded, and the second $\C^*$ action  completely fixes $x_4$, yielding $\C^2-\{x_1=x_3=0\}/\C^*$, which is now isomorphic to $\P^1(p-r,r)$. The same result can be established by the toric stack method: now $v_2=(1-r,-r)$, but  $\gcd(1-r,-r)=1$ for any $r$, and therefore $N_{\c D_2}\iso \Z$. Thus in this case there is no room for any non-trivial extra group actions, but the naive weighs $(p-r,r)$ could be changed due to the quotienting. But this doesn't happen.

The $Star$ of $\c D_2$ now consists of the cones $\c C_1$ and $\c C_2$. Let's see what  the lattice relation between $v_1$ and  $v_3$ modulo $v_2$ is. Assume that $a v_1+b v_3 \equiv 0 \mod v_2$. Let us look at the 2nd component: $-ar+b(p-r)\equiv 0 \mod r$, which is equivalent to $b\,p \equiv 0 \mod r$. If we assume that $\gcd(p,r)=1$, then $b$ must divide $r$. This then shows that the relation satisfied in $N_{\c D_2}\iso \Z$ by $v_1$ and  $v_3$ is the one inherited from $N=\Z^2$, i.e., $(p-r)v_1+rv_3=0$. We are not going to consider the case $\gcd(p,r)>1$ for reasons to be explained shortly.

Since the two $\C^*$ actions are independent of each other, $X_{p,q}$ is the direct product of $\P^1$ and $\P^1(p-r,r)$: $X_{p,q}\iso \P^1 \!\times \P^1(p-r,r)$. The same conclusion can be drawn from the intersection relations $D_1 \cdot D_3 =0$ $D_2 \cdot D_4$=0 and the linear equivalences (\ref{e:weights}). Let us summarize this in the following diagram, where $p_1$ and $p_2$ denote the projections:
\begin{equation}
\xymatrix{
  & X_{p,q}  \ar[dl]_{p_1}\ar[dr]^{p_2}&\\
   \P^1 & & \P^1(p-r,r) \,.}
\end{equation}
 
Both $\P^1$ and $\P^1(p-r,r)$  have strongly exceptional collections. One might hope that the {\em exterior tensor product} of the exceptional collections would be a strongly exceptional collection on $X_{p,q}$. On the other hand, if $p-r$ and $r$ have a common divisor, than the obvious strongly exceptional collection on $\P^1(p-r,r)$ is shorter than $p$ terms. This would result in an exceptional collection shorter than $2p$ terms on the associated $Y^{p,q}$ space. Using the techniques discussed in Section~\ref{sec:ypp}, we can produce a length $p$ exceptional collection on $\P^1(p-r,r)$, but that involves torsion Cartier divisors, which are harder to handle.\footnote{We hope to return to this issue in a future publication.} 
To avoid this, from now on we assume that $\mathbf{\gcd(p-r,r)=1}$, which is equivalent to $\mathbf{\gcd(p,r)=1}$.

The $p-q = 2r$ family contains the $Y^{p,p-2}$ spaces as a subfamily, with $r=1$. The conjectured collection of line bundles for the $Y^{p,p-2}$ spaces is  listed in Eq.~(\ref{trig:mor}). Using the linear equivalences (\ref{e:weights}) we can conveniently rewrite this collection 
\begin{equation}
\begin{split}
 &\O,\, \qquad  \O(D_3),\quad \qquad \O(2D_3),\qquad \ldots \qquad\,,\,\O((p-1)D_3)\,,\\
 &\O(D_2),\,  \O(D_2+D_3),\, \O(D_2+2D_3),\, \ldots\,,\O(D_2+(p-1)D_3)\,.
\end{split}
\end{equation}
But this collection is precisely, the suitably ordered, exterior tensor product of the exceptional collections on $\P^1$ and $\P^1(p-1,1)$, because of the isomorphisms: $\O(D_2) = p_1^* \O_{\P^1}(1)$ and $\O(D_3) = p_2^* \O_{\P^1(p-1,1)}(1)$. Note that in the $r=1$ case the second linear equivalence from (\ref{e:weights}) is simply $D_1 \sim (p-r)D_3$, and hence  $D_3$ is a minimal divisor. 

With a suitable modification the above statement in fact holds for all the $X_{p,q}$ spaces we considered in this section. For $r \geq 2$ the divisor $D_3$ is non-minimal, and we need to use a suitable replacement $D_3'$, where $\O(D_3')=p_2^* \O_{\P^1(p-r,r)}(1)$, whereas $D_3=r D_3'$. On the other hand $\O(D_2) = p_1^* \O_{\P^1}(1)$ still holds. With this notation we have
\begin{prop}
The collection of rank $1$ sheaves 
\[
\O_X,\, \O_X(D_3'),\, \O_X(2D_3'),\,  \ldots\,, \O_X((p-1)D_3')\,,
\]
\[ 
\O_X(D_2),\,  \O_X(D_2+D_3'),\, \O_X(D_2+2D_3'),\, \ldots\,,\O_X(D_2+(p-1)D_3')
\]
on the complex surface $X_{p,q}$ is strongly exceptional, for any $p$ and $q$ that satisfy $p-q = 2r$, for some $r\in \N$ and $\gcd(p,r)=1$.
\end{prop}
\begin{proof}
Once again the proof consists of a systematic computation of the  $\Ext$ groups  between  two elements in the collection, including any  element with itself. The techniques that we use are similar to the $Y^{p,p-1}$ case, but the details are very different: for $Y^{p,p-1}$ we were dealing with a blow-up, while here we have fibrations (although trivial).

We start by proving the fact that the collection of sheaves $\O_X,\, \O_X(D_3'),\, \O_X(2D_3'),\,  \ldots$, $\O_X((p-1)D_3')$ is strongly exceptional. The issue can be reduced to a problem on $Y=\P^1(p-r,r)$. We can again prove a Parseval-type identity
\begin{lemma}
${\mytext{Ext}}^{\,i}_X(p_2^*\O_Y(a),p_2^*\O_Y(b)) = {\mytext{Ext}}^{\,i}_Y(\O_Y(a),\O_Y(b))$, for all $a,b\in \Z$.
\end{lemma}
\begin{proof}[Proof of the Lemma]
Since the $p_2^*\O_Y(a)$'s are invertible sheaves, the problem reduces to computing $\H^i(X,p^*\c F)$, where $\c F$ is an invertible sheaf. To evaluate this, we can use the Leray spectral sequence:
\begin{equation}\label{e:lssa}
E_2^{p,q}=\H^p(Y,\,R^q p_{2*}\,p_2^*{\c  F})\Longrightarrow \H^{p+q}(X,\, p_2^*{\c  F})\,.
\end{equation}
Once again, by the projection formula $R^q p_{2*}\,p^*{\c  F} = R^q p_{2*}(p_2^*{\c  F}\otimes \O_X) = {\c  F}\otimes R^q p_{2*}\O_X$. To compute $R^q p_{2*}\O_X$ we use Grauert's theorem (Theorem 12.8 and Corollary 12.9 of \cite{Hartshorne:}), which shows that $R^q \pi_*\O_X = \O_Y$ for $q=0$ and $0$ otherwise.\footnote{Invoking Grauert's theorem is somewhat of an overkill, but it does the job in a heartbeat.} 
So the Leray spectral sequence (\ref{e:lss}) degenerates and $\H^{p}(X,\, \pi^*{\c  F}) \cong \H^{p}(Y,\, {\c  F}) $.
\end{proof}

Using the lemma, and the fact that the standard exceptional collection $\langle \O_Y,\O_Y(1),\ldots$, $\O_Y(p-1) \rangle$ on $Y ={\mathbb P}(p-r,r)$ is strong, we conclude immediately that  $\langle \O, \O(D_3'), \ldots\ ,\O((p-1)D_3')\rangle$ is a strong exceptional collection on $X$.

Next we deal with the exceptionality of $\O_X$ and $\O_X(\pm D_2)$. Considering the short exact sequence (SES)
\begin{equation}\label{e1a}
\ses{\O_X(-D_2)}{\O_X}{\O_{D_2} }\,,
\end{equation}
the associated cohomology long exact sequence (LES) implies that  $\H^i(X, \O_X(-D_2) ) = 0$, for  $i\geq 0$. 

Tensoring by  $\O_X(D_2)$ the SES (\ref{e1a}) gives another SES
\begin{equation} \label{e2a}
\ses{\O_X}{\O_X({D_2})}{\O_ {D_2}}\,,
\end{equation}
where we used the fact that the self--intersection $D_2 \cdot D_2 =0$.\footnote{The intersection ring of a toric stack is isomorphic to the intersection ring of the underlying variety (Lemma~5.1 of \cite{BorisovDM}). The orbifold Chow rings are very different though.}
The associated cohomology LES implies that $\H^0(X, \O_X({D_2}) ) = \C^2$, and $\H^i(X, \O_X({D_2}) ) = 0$ for  $i\geq 1$. 

To conclude, we need to compute the $\Ext$'s between the $\O_X(mD_3')$'s and the $\O_X(D_2+kD_3')$'s. This is equivalent to  computing $\H^i(X, \O_X(\pm D_2+mD_3'))$, for both positive and negative $n$. 

First, tensoring (\ref{e1a}) by $ \O_X(mD_3')$ gives
\begin{equation}\label{e11a}
\ses{\O_X(mD_3'-D_2)}{\O_X(mD_3')}{\O_{D_2}(mD_3') }\,.
\end{equation}
Let us remind ourselves that in the above equation $\O_{D_2}(mD_3')$ really means $j_* \O_{D_2}\otimes\O_X(mD_3')$, where $j\! :D_2 \hookrightarrow X$ is the embedding, and in fact $D_2 \iso Y \iso \P^1(p-r,r)$. Therefore $j_* \O_{D_2}\otimes\O_X(mD_3')= j_* \O_{\P^1(p-r,r)}(m)$. As a consequence 
\[
\H^i(X,  j_* \O_{\P^1(p-r,r)}(m))\iso \H^i(\P^1(p-r,r),   \O_{\P^1(p-r,r)}(m))  =  \H^i(Y,\O_Y(m))\, ,
\]
which implies that 
\begin{equation}\label{e11aa}
\H^i(X, \O_{D_2}(mD_3')) \iso  \H^i(Y,\O_Y(m))\, .
\end{equation}

On the other hand, using the lemma, $\H^i(X, \O_X(mD_3')=\H^i(Y,\O_Y(m))$. Combining this with Eq.~(\ref{e11aa}) we have: $\H^i(X, \O_X(mD_3')=\H^i(X, \O_{D_2}(mD_3'))$. The associated cohomology LES of (\ref{e11a}) then gives $\H^i(X, \O_X(mD_3'-D_2))=0$, for all $i$ and all $m$. 

Finally, tensoring (\ref{e2a}) by $ \O_X(mD_3')$ gives
\begin{equation}\label{e12a}
\ses{\O_X(mD_3')}{\O_X(mD_3'+{D_2})}{\O_{D_2}(mD_3') }\,.
\end{equation}
By the lemma $ \H^i(X, \O_X(mD_3'))= 0$ for all $i> 0$ and any $-(p-1)\leq m$. The cohomology LES then gives $ \H^i(X, \O_X(mD_3'+{D_2}))= 0$ for the same range of $i$ and $m$. Which completes the proof.
\end{proof}

\subsection{$X^{p,q}$}	\label{s:xpq1}


In Section~\ref{s:xpq} we already discussed the fact that the $X^{p,q}$ space is in a sense a blow-up of the $Y^{p,q-1}$ space at a smooth point. More precisely, if $X_{p,q}$ denotes the surface whose canonical sheaf is the 3-fold $X^{p,q}$, and similarly for $Y^{p,q-1}$, then $X_{p,q}$ is a blow-up of $Y_{p,q-1}$ at a smooth point. 

Given a collection on $Y_{p,q-1}$, one can immediately pull it back to $X_{p,q}$ using the blow-up map $\pi\! : X_{p,q}\to Y_{p,q-1}$, and try to augment it in a suitable way. We can in fact prove the following:

\begin{prop}
Let $\pi\! : X_{p,q}\to Y_{p,q-1}$ be the blow-up map. Given a strongly exceptional collection of invertible sheaves
$${\mathcal E} = \O_{Y_{p,q-1}}, E_2, E_3, \ldots E_n$$ 
on $Y_{p,q-1}$, the induced collection on $X_{p,q}$
$${\mathcal E}' = \O_{X_{p,q}},  \O_{X_{p,q}}(D_5), \pi^*E_2, \pi^*E_3, \ldots, \pi^* E_n$$ 
is also strongly exceptional.
\end{prop}

Whenever we know a strongly exceptional collection on $Y^{p,q-1}$, the proposition allows us to determine the gauge theory for the corresponding $X^{p,q}$ space.

\begin{proof}
Let  $E=D_5$ be the exceptional divisor of the blow-up $\pi\! : X_{p,q}\to Y_{p,q-1}$. Since we blew up a smooth point we expect that $E^2=-1$ and $E= \P^1$. These facts can be checked easily and explicitly by toric methods. First, there is a linear equivalence $D_2\sim D_4+D_5$. Second, the cones give the intersection products:  $D_2\cdot D_5=0$ and $D_4\cdot D_5=1$. Then
\begin{equation}
E\cdot E = D_5\cdot D_5=D_5\cdot (D_2 - D_4) = 0-1=-1\,.
\end{equation}
To show that $E= \P^1$ we use the BCS method. Since $v_5 = (p-q-2, p-q-1) $, and $\gcd(p-q-2, p-q-1)=1$, we have that $N_{D_5}=\Z$. The old relation $v_5 = v_4 + v_1$ now gives $v_4 + v_1\cong 0 \mod v_5$. Since $  v_1$ and $v_4 $ are the $Star$ of $v_5$ this shows that $E= \P^1$.

Now that $E^2=-1$ and $E= \P^1$ we can follow step by step the proof of Proposition~\ref{prop61}. To avoid repetition we will be sketchy on the details that are similar to the $Y^{p,p-1}$ case. Using Eq.~(\ref{e1}) and (\ref{e2}) an identical computation shows that $ \O_X$ and $ \O_X(E)$ form a strongly exceptional collection.\footnote{We drop the superscripts of $X_{p,q}$ and $Y_{p,q-1}$.} 

To deal with the tail $\langle \pi^*E_2, \pi^*E_3, \ldots, \pi^* E_n \rangle$ we can use once again Lemma~\ref{prop6lemma}, and the assumption that ${\mathcal E} = \langle \O_{Y_{p,q-1}}, E_2, E_3, \ldots E_n\rangle$ is strongly exceptional on $Y$.

Finally we have to glue the two strong exceptional collections on $X$: $\langle \O_X, \O_X(E)\rangle$ and $\langle \pi^*E_2, \pi^*E_3, \ldots, \pi^* E_n \rangle$. This boils down to computing $\H^i(X, \pi^*E_m(\pm E))$, for $m\in\Z$, both positive and negative.

Let's do $\H^i(X, \pi^*E_m(- E))$ first. Tensoring (\ref{e1}) with $\pi^*E_m $ gives
\begin{equation}\label{e11z}
\ses{\pi^*E_m(-E)}{\pi^*E_m}{\O_E \otimes \pi^*E_m}\,.
\end{equation}
Now remember that $\O_E \otimes \pi^*E_m$ is in fact $j_*\O_E \otimes \pi^*E_m$, where $j\! : E \hookrightarrow X$ is the embedding. Using the projection formula 
\begin{equation}\label{e:projft}
j_*\O_E \otimes \pi^*E_m = j_*(\O_E \otimes j^*\pi^*E_m)= j_*(\O_E \otimes (\pi\comp j)^*E_m)\,.
\end{equation}
But the map $\pi\comp j\! : E \hookrightarrow Y$ projects $E$ to a point, the one that was blown up. By assumption the $E_m$'s are invertible sheaves, and hence $ (\pi\comp j)^*E_m=\O_E$.

From the  cohomology LES  associated  to (\ref{e11z}) we have that $\dim \H^0(X, \pi^*E_m(- E))=$  $ \dim \H^0(X, \pi^*E_m)-1$,
while $ \H^i(X, \pi^*E_m(- E))= \dim \H^i(X, \pi^*E_m)$ for $i>0$. Using Lemma~\ref{prop6lemma}: $\H^i(X, \pi^*E_m)=\H^i(Y, E_m)$. The strong exceptional property of the collection on $Y$ guarantees that the latter one is $0$ for any $i>0$.

Finally, tensoring (\ref{e2}) with $\pi^*E_m^{\, \roof }$, the dual of $\pi^*E_m $, gives
\begin{equation}\label{e12z}
\ses{\pi^*E_m^{\, \roof}}{\pi^*E_m^{\, \roof}(E)}{\O_E(-1)\otimes \pi^*E_m^{\, \roof} }\,.
\end{equation}
A computation identical to (\ref{e:projft}) shows that $j_*\O_E(-1)\otimes \pi^*E_m^{\, \roof} =j_*\O_E(-1)$. The cohomology LES  associated  to (\ref{e12z}) then completes the proof.
\end{proof}

\section{Exceptional Collection Conjecture}	\label{s:exconj}

Given a singular toric surface generated by four rays $v_1$, $v_2$, $v_3$ and $v_4$
such that the hull of the four rays is a convex quadrilateral, we conjecture a strongly
exceptional collection of line bundles.  
To motivate this conjecture, we start with
the Calabi-Yau cone, promoting our 2-dimensional vectors $v_i$ to coplanar three vectors $V_i = (1, v_i)$.
While the $v_i$ will in general satisfy two relations, the $V_i$ will in general satisfy
only one such relation which we take to have the form
\be
a V_1 - c V_2 + b V_3 - d V_4 = 0 \,,
\label{CYrel}
\ee
where we choose the integers to satisfy the inequalities $0 < a \leq c \leq d  \leq b$.  
The Calabi-Yau condition means that $a+b = c+d$.  (Note that the $v_i$ also satisfy this relation
(\ref{CYrel}).)  
From these weights $(a,-c,b,-d)$ we will construct a quiver gauge theory.  From the quiver,
we will extract a collection of line bundles which we believe form a complete and strongly exceptional collection.

We insist that ${\mbox{gcd}}(a,b,c,d)=1$ or our method will give
several copies of the quiver for $(a,b,c,d)/\mbox{gcd}(a,b,c,d)$.  
More precisely,
we would like there to be no torsion divisors in the singular surface and so take
the $v_i$ to span the ${\mathbb Z}^2$ lattice.  

To construct the quiver, we rely on the relation between bifundamental fields in the
gauge theory and divisors.  For these toric varieties the simplest divisors are the ones
invariant under the torus-action, which are in one-to-one correspondence with the rays
of the quiver.  For each $V_i$, we have a T-invariant divisor $D_i$.    
From \cite{HananyLpqr} the number of bifundamental fields corresponding to $D_i$
which appear in 
the quiver gauge theory is equal to the determinant $\langle V_{i-1}, V_i, V_{i+1} \rangle$ 
where we
take $V_{5}=V_1$ and $V_{0} = V_4$.  The proof comes from considering how 
many torsion line bundles can be constructed from each such divisor.

If we choose the $v_i$ to be
\be
v_1 = (1,0) \; , \; \; \; v_2 = (ak,b) \; , \; \; \; v_3 = (-al,c) \; , \; \; \; v_4 = (0,0)
\ee
where $ck+bl=1$, then we find that there are $b$ bifundamental fields of type $D_1$, 
$d$ bifundamentals of type $D_2$, $a$ bifundamentals of type $D_3$,
and $c$ bifundamentals of type $D_4$.

From our exceptional collection techniques we expect that each node in the quiver can be identified with a line bundle
on the Calabi-Yau cone $X$.  A line bundle on $X$ will in general take the form $\calo(\sum k_i D_i) $. Since there are only four rays generating $X$, all the $D_i$ are linearly equivalent and we can write this line bundle in an equally descriptive way as 
\be
\calo(\sum k_i D_i) \cong \calo(a k_1 - c k_2 + b k_3 - d k_4) \ .
\ee
We can thus think of a collection of line bundles $\cale = (E_1, E_2, \ldots, E_n)$
as a set of integers, $\cale = (m_1, m_2, \ldots, m_n)$ (up to torsion).

To have a complete exceptional collection, we need one line bundle for each type of 
D-brane.  This number of objects is just the Euler character $\chi$ of the Calabi-Yau cone,
which is also the number of triangles in a complete triangulation of the base of
the toric cone \cite{Fulton}.  In our case, $\chi = a+b=c+d$ \cite{HananyLpqr}.

The differences of line bundles in the exceptional collection can be identified with
sums of divisors.  If the difference is one of the $D_i$, then we can place
a bifundamental field between those two nodes because there is no
simpler way of expressing $D_i$.  Thus, our exceptional collection
will consist of a set of integers $\cale = (m_1, m_2, \ldots, m_{a+b})$, which satisfy the following condition: there are $b$ pairs $i,j$, such that $m_j - m_i = a$, 
$d$ pairs such that $m_j - m_i = c$, $a$ pairs such that $m_j - m_i = b$, and
$c$ pairs such that $m_j - m_i = d$.  There are in general many solutions
to these constraints which we expect to be related by Seiberg duality.  
However, one solution which always seems to work is the following.

Consider the set 
\be
\cale = (N+1, N+2, \ldots, N+a+b) \ .
\label{skeleton}
\ee
We draw all possible arrows from $E_i$ to $E_{i+a}$, from $E_i$ to $E_{i+b}$, from
$E_{i+c}$ to $E_i$, and from $E_{i+d}$ to $E_i$.
We are almost done, but there is a subtlety.

There may also be some bifundamental fields of the form $D_i+D_j$. 
Consider $D_3+D_4$ first.  
The charge of this divisor
is $b-d$.  If we have two nodes $m_{i+b-d}$ and $m_i$, then the difference must
correspond to the divisor $D_3+D_4$.  However, it is not always true that
we can think of this map as composed individually of a $D_3$ map from
$m_j$ to $m_{j+b}$ and a 
$D_4$ map from $m_k$ to $m_{k-d}$.  We find that for $a < i \leq d$, no such
combination exists.  So we need to add $d-a$ bifundamental fields corresponding
to the divisor $D_3+D_4$ for $i$ in this range.  
A similar situation exists for $D_2+D_3$.  Here
the charge is $b-c$.  Now there will be no combination of maps that sum to
$D_2+D_3$ for $a < i \leq c$.   In other words, we must add $c-a$ fields of type
$D_2+D_3$ that begin on nodes $i$ for $a < i \leq c$.
We don't need to draw any arrows for $D_1+D_2$ or for $D_1+D_4$ because
for our $\cale$, these divisors can be produced by combining maps corresponding
to the individual $D_i$.  A similar situation holds for $D_1+D_3$, $D_2+D_4$, and
the $2D_i$.

Let us do an example.  Consider $L^{2,6,3}$ which has the single relation
\be
2 V_1 - 3 V_2 + 6 V_3 - 5 V_4 = 0 \ .
\ee
There will be eight nodes in the quiver ($6+2=8$), and eight objects in the collection
\be
\cale = (E_1, E_2, \ldots, E_8) \ .
\ee
We first draw in 
the arrows corresponding to the $D_i$.  There will be six arrows from $E_i$ to $E_{i+2}$
and two arrows from $E_i$ to $E_{i+6}$.  In the opposite direction, there will be
five arrows from $E_i$ to $E_{i-3}$ and three arrows from $E_i$ to $E_{i-5}$.  Finally
we add in the arrows corresponding to the missing $D_2+D_3$ and $D_3+D_4$.
There will be three arrows corresponding to $D_3+D_4$, one from $E_3$ to $E_4$, one
from $E_4$ to $E_5$ and one from $E_5$ to $E_6$.  There will be one arrow 
corresponding to $D_2+D_3$ from $E_3$ to $E_6$.  This quiver, up to a 
reordering of the nodes, is exactly the same
as Fig.~9 of \cite{HananyLpqr}.

\FIGURE{
\epsfig{file=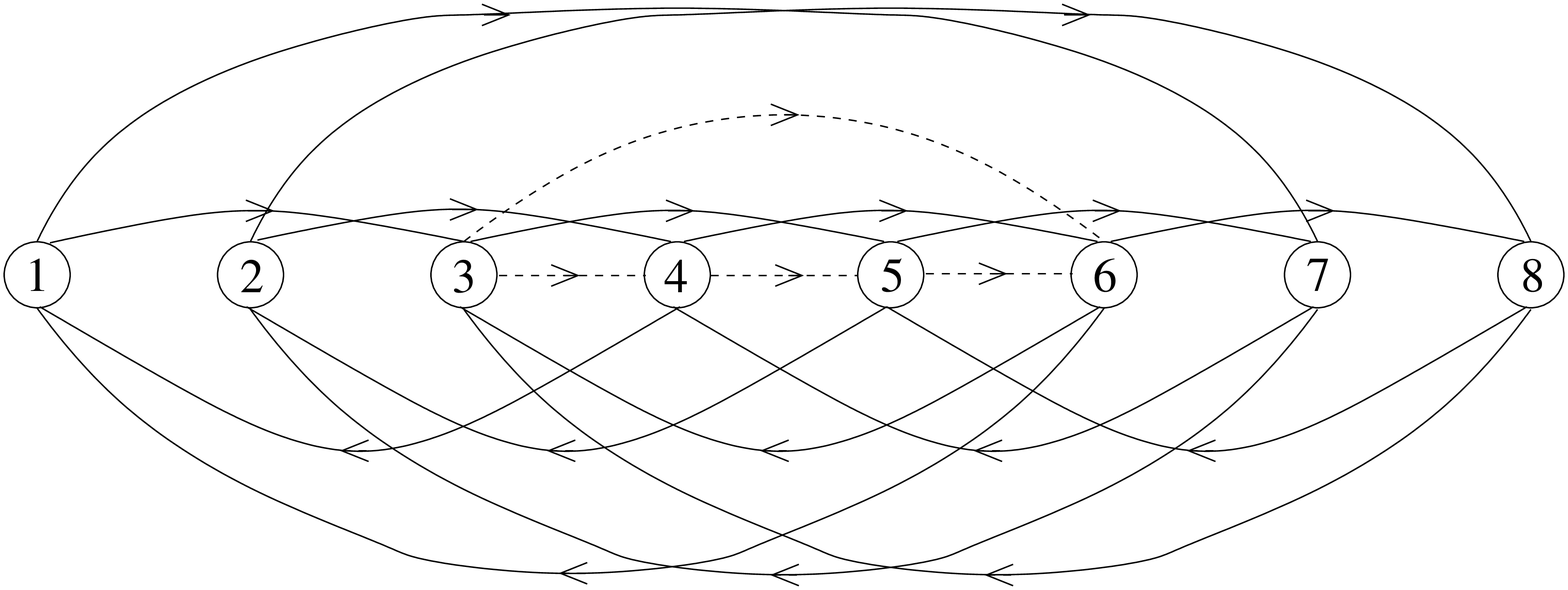, width=2.9in}
\caption{The gauge theory quiver for $L^{2,6,3}$.  The dashed arrows
correspond to the composite divisors $D_3+D_4$ and $D_2+D_3$.}
\label{L263}
}


Thus we have  a simple recipe for computing a quiver given a toric Calabi-Yau cone generated by
four rays.  We also have the skeleton of an exceptional collection on a toric surface 
corresponding to the $v_i$.  
While the cone has only one relation, the toric surface will have two relations, one of which
continues to be $(a,-c,b,-d)$.
The toric surface, because it has an extra relation, will have one
additional linearly independent divisor.  Thus, one integer $m_i$ is no longer enough to 
specify uniquely a line bundle.  We need two integers: the set  
\be
\begin{split}
\calo_V(&N+1, n_1), \calo_V(N+2, n_2), \ldots \\
&\ldots, \calo_V(N+a+b, n_{a+b})
\end{split}
\label{skeletonV}
\ee  
appropriately reordered for some set of integers $n_i$ and $N$, should be a complete, 
strongly exceptional collection on our toric surface.  

Motivated by the smooth del Pezzo case where the sheaves are ordered by
their slope (as reviewed in section \ref{sec:ordering}), 
we would like to choose $n_i$ to be the degree of the line bundle.
Having found a set of $n_i$ that are sandwiched between 0 and $K^2$,
we conjecture that the collection will be strongly exceptional ordered by 
the $n_i$.

Given the quiver and the identification of arrows in the quiver with toric divisors $D_i$,
we can momentarily disregard this skeleton collection, and build a new
collection from scratch.  Pick any node of the quiver and associate to it $\calo_V$.
Working outward from $\calo_V$, 
given two nodes $i$ and $j$ and an arrow from $i$ to $j$ corresponding to
a divisor $D$, 
we associate line 
bundles $\calo_V(C_i)$ and $\calo_V(C_j)$ to the nodes such that $C_i + D = C_j$.
As there are many loops in the quiver, this procedure is not uniquely determined.
Also note that because of these loops, 
we will always find nodes connected by an arrow such that
$C_i + D \neq C_j$.  However in this case, if we let
\be
C_i + D - C_j = \sum k_i D_i
\ee  
then, $a k_1 - c k_2 + b k_3 - d k_4 = 0$.  In other words, as we can see from
our skeleton collection, the discrepancy will be trivial when lifted to the Calabi-Yau cone.

At this point, we need to check whether all our line bundles have degree bounded
between 0 and $K^2$.  We take a brief detour to explain how to calculate
$-K \cdot D$ from the fan. What we have to do is toric intersection theory on an orbifold, and is well known \cite{Fulton}.  Given $D_i$ and $D_j$, $i \neq j$, if $v_i$ and $v_j$ are not rays of the same maximal cone, then $D_i \cdot D_j = 0$.  On the other hand, if $v_i$ and $v_j$ do belong to a maximal cone, then  $D_i \cdot D_j$ is the reciprocal of the index of the sub-lattice generated by $v_i$ and $v_j$ in $\Z^2$. The index can be conveniently computed using a two-by-two determinant:
\[
D_i \cdot D_{i+1} = \frac{1}{|  v_i, v_{i+1}  |}\,.
\]
For a fan generated by
$n$ rays, these $n(n-1)/2$ relations generally overdetermine the intersection matrix.
The remaining $D_i^2$ can be determined from linear equivalence, picking two of the
divisors and expressing them as linear combinations of the remaining $n-2$.  

In this not uniquely determined procedure for labeling the nodes of the quiver, we search
out the line bundle with the smallest and the line bundle with largest degree.  If the difference
in degree is more than $K^2$, then we take the largest degree line bundle and add $K$.
We continue this procedure until all the line bundles are sandwiched between 0 and $K^2$.

Again, consider the example $L^{2,6,3}$.  
We take the fan of the toric surface to be
\be
 v_1 = (1,1)\; , \; \; \; v_2 = (-1,0) \; , \; \; \; v_3 = (0,-2) \; , \; \; \;
v_4 = (1,-2) \ .
\ee
We will call $E_3 = \calo_V$.  Then we find, 
in the wrong order suggested by the skeleton collection,
\be
\begin{split}
\cale = & \calo_V(3D_3+4D_4), \calo_V(4D_3+5D_4), \calo_V,	\\
&\calo_V(D_3+D_4), \calo_V(2D_3+2D_4),  \ldots, \calo_V(5D_3+5D_4) \, .
\end{split}
\ee
From the fan, we find that 
\be
D_3 \cdot D_4 = \frac{1}{2} \; , \; \; \;
D_3^2 = -\frac{1}{2} \; , \; \; \;
D_4^2 = -\frac{1}{3} \ ,
\ee
and, using linear equivalence, $-K = 5D_3 + 6D_4$.  Thus, 
$-K \cdot (a_3 D_3 + a_4 D_4) = (a_3+a_4)/2$.  Thus, 
the correct order should be
\[
\cale = \calo_V,  \calo_V(D_3+D_4), \calo_V(2D_3+2D_4), \calo_V(3D_3+3D_4),
\calo_V(3D_3+4D_4), 
\]
\be 
\qquad \qquad
\calo_V (4D_3 + 4D_4), \calo_V(4D_3+5D_4), \calo_V(5D_3 + 5D_4)  \ .
\ee
This collection is indeed sandwiched between 0 and $K^2 = 11/2$.
In labeling the nodes, we could easily have chosen $\calo_V(2D_1+2D_2) \sim \calo_V(8D_3 + 10D_4)$ instead of $\calo_V(3D_3 + 4D_4)$.  After calculating the degrees, we would
have found that the collection is not sandwiched between 0 and $K^2$ and would have
known to add $K$ to $8D_3 + 10 D_4$.

Indeed, our computer routines verified that this collection is strongly exceptional
in the order presented.  We can then complete the circle and 
check that this collection also generates the correct quiver.
The computer routines yield
\be
 S^{-1} = 
\left(
\begin{array}{rrrrrrrr}
1 & -1 & -1 & 0 & 1 & 0 & 0 & 1 \\
0 & 1 & -1 & -1 & -1 & 1 & 1 & 0\\
0 & 0 & 1 & -1 & 0 & -1 & -1 & 1 \\
0 & 0 & 0 & 1 & -1 & 0 & 0 & -1  \\
0 & 0 & 0 & 0 & 1 & -1 & 0 & 0 \\
0 & 0 & 0 & 0 & 0 & 1 & -1 & 0\\
0 & 0 & 0 & 0 & 0 & 0 & 1 & -1 \\
0 & 0 & 0 & 0 & 0 & 0 & 0 & 1
\end{array}
\right) \ .
\ee
This matrix gives rise to precisely
the quiver discussed above.  There is one subtlety however.  The quiver
we discussed has a bidirectional arrow between what we have labeled nodes
1 and 4.  We find $S^{-1}_{14} = 0$ and would have to study
explicitly the inverse sheaves to detect these two maps from the collection.

Before concluding this section, we consider one more example, $Y^{3,1}$.
As above, we take the $v_i$ to satisfy the relation $4v_1 -3v_2 + 2 v_3 -3v_4$. 
Before, we considered two distinct strongly exceptional collections for this
space.  There was the collection that gave rise to a quiver 
with two single impurities (\ref{Yp0conj}):
\be
\cale = \calo_V, \calo_V(D_4), \calo_V(D_3+D_4), \calo_V(D_3+2D_4),
\calo_V(2D_3+2D_4), \calo_V(2D_3+3D_4) \ .
\ee
Lifting this collection to the Calabi-Yau cone, we find
\be
\cale = (0,-3,-1,-4,-2,-5)
\ee
which is precisely of the form (\ref{skeleton}).  We also found a Seiberg dual quiver
with a single double impurity (\ref{Y31double}):
\be
\cale = \calo_V, \calo_V(D_3), \calo_V(2D_3), \calo_V(D_4), \calo_V(D_3+D_4), 
\calo_V(2D_3+D_4)  \ .
\ee
Lifting this collection to the cone yields
\be
\cale = (0,2,4,-3,-1,1) \ 
\ee
which is not of the form (\ref{skeleton}).  Nevertheless, one can check that there
are still 2 fields of type $D_3$, 4 fields of type $D_1$, and 3 fields of both
type $D_2$ and $D_4$, as required.

\section{Discussion}

What we have done in this paper is nontrivial and surprising.
Through a combination of guesswork and deduction, we have found 
strongly exceptional collections on a large number of singular toric
surfaces.  Using some basic results in toric geometry and some more sophisticated
results in local cohomology, we were able to prove that these collections
were indeed strongly exceptional and from these collections derive quiver
gauge theories for the D-branes probing the corresponding Calabi-Yau singularities.

A priori there were no guarantees this program would succeed.  Although
exceptional collections were known to exist on these singular surfaces \cite{Kawamata}, 
it was not clear that the more physical strongly exceptional collections could be found.
It was also unclear that these singular surfaces -- the result of only partially resolving
the Calabi-Yau singularity -- would contain the gauge theory information.
It was not even clear that the mathematical technology existed to manipulate these
collections once obtained.  Indeed, some of the requisite stacky techniques were only
developed in the last few years.

Although this paper is a good beginning, much remains to be done.  
\begin{enumerate}
\item
We need a general algorithm for converting a toric diagram into an exceptional
collection.  The techniques presented in this paper are quite powerful, and
eventually we hope the exceptional collection method will be both the most
rigorous and fastest way of converting geometry into gauge theory.
\item
These ${\mathcal N}=1$ supersymmetric gauge theories are specified
by both a quiver and a superpotential.  While we gave quivers,
we did not derive superpotentials in this paper.  Techniques exist
for computing the superpotential from an exceptional collection
in the smooth case \cite{Wijnholt, Aspinwall-Katz, Aspinwall-Fidkowski},
and we expect that these techniques generalize to the singular case as well.
\item
We said very little about Seiberg duality.
While we emphasized at the beginning that exceptional collections
can produce conformal 
quiver gauge theories where the ranks of the gauge groups
are not all equal, we only presented collections of line bundles
where the ranks were equal.  Seiberg duality 
will generically
produce these higher rank gauge theories.  
Given the collections here, it should be straightforward to
explore these equivalence classes of Seiberg dual theories.
\item
Seiberg duality is defined as sheaf mutations only for
well split nodes \cite{Herzog2}.  For the smooth del Pezzo
case, this definition was ample because the physical
exceptional collections all appear to be well-split.
One burning
question is how to define Seiberg duality for the ill-split
$L^{p,q,r}$ examples discussed in the text.
\item
Although we asserted that the bifundamental fields in the gauge
theory come from the maps between the inverse collection,
we never computed this inverse collection explicitly, just $S^{-1}$.  
The inverse collection, also often called a collection of
fractional branes, lives in the derived category.  
In the smooth
case, the fractional branes were one--term complexes.  In
the singular case one expects more complicated complexes. This has already been observed for the $\C^3/\Z_5$ orbifold \cite{en:fracC3}, and it is true even in the  $\C^2/\Z_n$ case for $n >3 $ \cite{en:fracC2}.
\item
There are a number of mathematical results for 
exceptional collections on del Pezzos that one would
ideally hope to establish rigorously for the singular cases.
One such result involves the existence of a helix.
In the smooth case, one can prove that right mutating
the $E_1$ in $\cale$ to the end of the collection is equivalent
to tensoring $E_1$ by the anti-canonical class.  For the examples
studied in the text this same result appears to hold in the singular
case, but we lack a proof at the moment.
\end{enumerate}

\section*{Acknowledgments}

We would like to thank Aaron Bergman, Emanuel Diaconescu, Charles Doran, Pavel Kovtun, 
James McKernan, Greg Moore, David Morrison,
James Parson, 
Sam Pinansky, 
Hal Schenck, Eric Sharpe and  Mike Stillman for discussions.
The research of C.~P.~H. was supported in part by the NSF under Grant No.~PHY99-07949. R.~L.~K. was supported in part by a DOE grant DE-FG02-96ER40949.

\newpage
\begin{appendix}

\section{Computing  sheaf valued toric cohomology}    \label{a:intro}

There are several useful ways to construct a toric variety $X$ given its fan $\Delta$. We will mention only two of them.\footnote{For a thorough discussion see \cite{Cox:Katz}, and references therein.} 
Probably the most widely known method is to use the group algebra $\C[\sigma^{\roof}]$ of every dual cone $\sigma^{\roof}$, and patch the associated affine schemes together, as explained for example in \cite{Fulton}. In the case of projective spaces this construction is the analog of the affine coordinate charts.
 
For projective spaces on the other hand we also have the very efficient homogeneous coordinates at our disposal, and one can ask if there is an analog of this for toric varieties. This question was answered positively by Cox in 1995 \cite{Cox:HoloQout}. Cox's construction is in terms of a GIT quotient, and is usually referred to as the {\em holomorphic quotient}. His construction generalizes the $\C^{n+1}-\{0\}/\C^*$ representation of $\C\P^n$, replacing $\{0\}$ with a suitable subset and assigning certain weights to the different $\C^*$ generators. We will review elements of this construction shortly.

Our main interest is in computing sheaf cohomology groups for locally free sheaves, and in particular invertible ones, i.e., line bundles. The two toric variety constructions mentioned previously give two very different ways to achieve this. Computing $\H^0(X,\O(D))$ in the ``old'' affine-patching way boils down to counting lattice points inside a polytope associated to $D$. Higher cohomologies are harder to compute, but there is a well-known, if not very attractive, method \cite{Fulton}: one has to compute relative cohomology groups of simplicial topological spaces. 

Another method to compute sheaf cohomology for torus equivariant bundles is popularized in \cite{Knutson:Sharpe}, based on the work of Klyachko \cite{Klyachko}, and uses flags of vector spaces. For simple examples this is indeed efficient, but we found little practical use for it in our context. 

Therefore it is natural to ask whether a cohomology computation based on the holomorphic quotient construction would be more convenient for practical purposes. This would have obvious advantages which extend beyond the scope of this paper, for example to the $(0,2)$ gauged linear sigma models \cite{Witten:GLSM,Distler:1996tj}. Partly motivated by the authors of \cite{Distler:1996tj}, Eisenbud, Mustata and Stillman tackled the problem of computing $\H^i(X,\O(D))$ in a uniform way using the toric homogeneous coordinate ring of Cox \cite{ToricCoh}. Since this work is virtually unknown in the physics community, we will review their approach in more detail. For the convenience of the reader we also follow their notation.

\subsection{Gradings}	\label{a:app1}

We begin with Cox's homogeneous coordinate ring. Since this is well known, we only focus on some aspects that are relevant for the cohomology. Let $\Delta$ be the fan in $\Z^d$ corresponding to the toric variety $X$, and assume that $\Delta$ has $n$ edges (one-dimensional cones). To every  edge of $\Delta$ one associates a variable, and we get a polynomial ring $S := \C[x_1,\dots,x_n]$ . $S$ has two obvious gradings: 
\begin{enumerate}
\item $\Z$-grading: every variable is assigned degree $1$;
\item $\Z^n$-grading: $x_i$ has degree $1$ under the $i$th $\Z$, and $0$ under all other  $\Z$'s, i.e., $\deg x_i =(0,\dots,\overset{i{\rm th}}1,\dots,0)$.
\end{enumerate}
Cox introduces a new grading for $S$ which is finer than the $\Z$-grading but coarser than the $\Z^n$-grading. Consider the matrix $\rho$, whose rows are the coordinates of the first integral points of the $n$ edges of $\Delta$.\footnote{Although the original paper \cite{Cox:HoloQout} uses the minimal generators of the ray, the construction extends naturally to non-minimal generators.} 
After a choice of basis this matrix represents a linear map, and the torus-invariant divisor classes correspond to the elements of the cokernel $\D$ \cite{Fulton}, yielding an exact sequence
\begin{equation}\label{e:zw1}
\xymatrix@1{ \Z^d \ar[r] ^\rho& \Z^n \ar[r]^\phi &  \D \ar[r] & 0}\,.
\end{equation}
Now we introduce the grading that is coarser than the $\Z^n$-grading: the $\D$-grading. 

For concreteness let us consider the mother of all examples, the projective space: $\P^n$. There are $n+1$ vertices 
$v_1=(1,\ldots,0) , \ldots ,v_n=(0,\ldots,1),v_{n+1}=(-1,\ldots,-1)\in \Z^n$ , and the fan has the obvious $n+1$ cones. The matrix $\rho$ in (\ref{e:zw1}) in this case becomes
\begin{equation}
\rho=
\left(\!\!
\begin{array}{rrrr}
1&0&\cdots&0\\
0&1&\cdots&0\\
\vdots&\vdots&\ddots&\vdots\\
0&0&\cdots&1\\
-1&-1&\cdots&-1
\end{array}
\right)
\end{equation}
with cokernel isomorphic to  $\Z$. For future convenience call the generator of the cokernel $\Upsilon$.

Any $\Z^n$-graded module $N$ (and in particular $S$ itself) can be regarded as a $\D$-graded module by setting 
\begin{equation}
N_\delta=\underset {p\in\phi^{-1}\delta}\bigoplus N_p \quad	\mbox{for each} \quad \delta\in\D\,.
\end{equation}
In other words the  ring $S := \C[x_1,\dots,x_n]$ is graded by the divisor class group,  and the degree of $x^D=\prod x_i^{a_i}$ equals
\begin{equation}
\deg(x^D) = [\sum a_i D_i] \in \operatorname{DivCl}(X)\,,
\end{equation} 
where $D_i$ is the divisor associated to the $i$th edge. For our $\P^n$ we simply have $\deg(x_i) = 1\cdot \Upsilon$, for all $ i=1,\ldots,n+1$. This is the usual assignment: the homogenous coordinates of $\P^n$ all have degree $1$.

The {\it homogeneous coordinate ring\/} of $X$ is the polynomial ring $S$ together with the $\D$-grading and the {\it irrelevant ideal\/}
\begin{equation}
B\ =\  \langle \{\prod_{v_i \not\in \sigma} x_i \mid \sigma \in \Delta\} \rangle\,,
\end{equation}
where $\sigma$ ranges over the cones in the fan $\Delta$. As an aside let us note that the zero set of the ideal $B$, ${\mathbb V}(B)$, is the exceptional set one discards in the GIT quotient and $\D$ gives rise to the $\C^*$ actions \cite{Cox:HoloQout}. For our example $\P^n$, $B=\langle x_1 ,x_2, \ldots , x_{n+1}\rangle$ and ${\mathbb V}(B)=(0,0,\ldots,0)$, as expected.

The graded pieces of $S$ have a well-known cohomological interpretation \cite{Cox:HoloQout,Cox:Katz}. If a Weil divisor $\D$ is not Cartier, then it determines a reflexive sheaf $\O_X(D)$, rather than a line bundle. Regardless of this, for any   Weil divisor $\D$ we have an isomorphism:\footnote{$S_\delta$ is the traditional way of denoting the collection of elements of degree $\delta$.}
\begin{equation}\label{e:d1} 
\H^0(X,\O_X(D))  = S_\delta\,, \quad \mbox{where $\delta=[D] \in \operatorname{DivCl}(X)$}\,.
\end{equation}
Therefore the $0$th cohomology of an invertible sheaf can be computed from the knowledge of the $\D$-grading, which in turn is a purely algebraic Hilbert-function  computation, and can be efficiently handled by ``the machines''. 

One observes the conceptual similarity in the computation of $\H^0$ between counting lattice points inside a polytope and counting monomials of a given degree. So what is the analog of the relative cohomology in the homogeneous coordinate context? The work of \cite{ToricCoh} extends this analogy to higher cohomologies and general toric sheaves (not only line bundles) in terms of Grothendieck's \textit{local cohomology}. But before we can present this, we need some more background material.

Given a $\D$-graded $S$-module $P$, Cox \cite{Cox:HoloQout} constructs a quasi-coherent sheaf $\tilde P$ on $X$ by localizing as in the case of projective space. This is usually called the Serre functor. Coherent sheaves come from finitely generated modules:
\begin{thm}{\rm(Theorem 2.1 in \cite{ToricCoh})}
Every coherent $\O_X$-module may be written as $\tilde{P}$ for a finitely generated $\D$-graded $S$-module $P$.
\end{thm}

For any $\D$-graded $S$-module $P$ and any $\delta\in \D$ it is customary to define the \textit{shifted} graded module $P(\delta)$, with the grading 
\begin{equation}
P(\delta)_\epsilon=P_{\delta+\epsilon}\,, \quad \mbox{for all $\epsilon\in \D$}\,.
\end{equation}
We introduce the following notation
\begin{equation}
\H^i_{tot}(X,\tilde P)= \underset {\delta\in \D}\bigoplus  \H^i(X,\widetilde {P(\delta)}).
\end{equation} 
In particular
\begin{equation}
\H^i_{tot}(X,\O_X)= \underset {\delta\in \D}\bigoplus  \H^i(X,\O_X(\delta))
\end{equation} 
contains all the cohomology groups that we want. Note that by the above discussion (\ref{e:d1})  immediately implies that $\H^0_{tot}(X,\O_X)\iso S$. 

The advantage of working with $H^i_{tot}(X,\tilde{P})$ is that it can be computed in terms of local cohomology, which in turn can be obtained as a limit of Ext modules. The first relevant result for us in this direction is an extension to toric varieties of a well-known projective result (see, e.g., Appendix A4 of \cite{Eisenbud}):
\begin{thm}{\rm (Prop. 2.3 in \cite{ToricCoh})}
Let $P$ be a $\D$-graded $S$-module, $\tilde{P}$ be the corresponding quasi-coherent sheaf on $X$, and let $\H^i_B(P)$ denote the local cohomology module with support in $B$. Then we have:

(a) For $i \geq 1$, there is an isomorphism of graded $S$-modules
\begin{equation}
\H^i_{tot}(X,\tilde{P}) \, \cong \, \H^{i+1}_B(P)\,.
\end{equation} 

(b) There is an exact sequence of graded $S$-modules
\begin{equation}
\xymatrix@1{0 \ar[r] & \H^0_B(P) \ar[r] & P \ar[r]  & \H^0_{tot}(X,\tilde{P}) \ar[r] & \H^1_B(P)  \ar[r] & 0}\,.
\end{equation} 
\end{thm}

Let us spend a few moments elucidating the definition and intuition behind local cohomology. This will then illuminate the main result of \cite{ToricCoh}, which we use for the explicit computations. For a concise introduction to local cohomology one can consult \cite{Eisenbud}, while \cite{LocalCohomology} provides a detailed treatment. 

\subsection{Local Cohomology}

Local cohomology was introduced by Grothendieck as an algebraic analog of the topological relative cohomology \cite{Hartshorne:Local}. One can define local cohomology using a \v Cech complex, but instead we follow the abstract derived functor approach. To avoid confusion, let us add that local cohomology is in fact more general than sheaf cohomology, at least for projective
schemes. So the name is a misnomer.

Let $A$ be a ring, $I$ be an ideal of $A$, and $M$ be an $A$-module. The {\em $0$th local cohomology module of $M$ with support in $I$} is defined as the set of all elements of $M$ that are annihilated by some power of $I$:
\begin{equation}\label{e:lc1} 
\Gamma^0_I(M) = \underset{n\in \mathbb N} \cup (0:_M I^n)\,.
\end{equation} 
Here we used the standard notation $(0:_M I^n) = \{m\in M| \ am=0 \ \mbox{for all $a\in I^n$}\}$. First, one observes the inclusion $(0:_M I^n)\subset (0:_M I^{n+1})$. Therefore the union in (\ref{e:lc1}) is in fact a direct limit. Second, there is an isomorphism of $A$-modules $(0:_M I^n) \iso \Hom_A(A/I^n, M)$.\footnote{Given $m\in (0:_M I^n)$ one can define an $A$-module homomorphism $\varphi_m\!: A/I^n\to M$ by sending the class $a+I^n$ to $am$. Conversely, given an $A$-module homomorphism $\phi\!: A/I^n\to M$  one has $\phi(1)\in (0:_M I^n)$.}
Since $I$ is an ideal,  $I^{n+1}\subset I^n$, and thus $ \Hom_A(A/I^n, M)\subset  \Hom_A(A/I^{n+1}, M)$, exhibiting the earlier observed filtration. All this can be summarized in the form
\begin{equation}\label{e:lc2} 
\Gamma^0_I(M) = \lim_{n\to \infty} \Hom_A(A/I^n, M)\,.
\end{equation} 

There is a nice geometric interpretation of local cohomology. If we think of the elements of  $M$ as global sections of the sheaf $\tilde{M}$ on $\Spec A$, then the elements of  $\Gamma^0_I(M) $ are the sections with support on the closed subscheme $\Spec A/I$. 

One can show that $\Gamma^0_I(-)$ is a {\em left} exact functor. This allows us to define the higher local cohomology modules as its {\em right}-derived functors:\footnote{One uses an  injective resolution for $M$.}
\begin{equation}
\H^i_I(M) = R^i\Gamma^0_I(M)\,.
\end{equation} 
The useful thing for us is the fact that the representation (\ref{e:lc2}) has a nice extension:
\begin{equation}\label{e:lc3} 
\H^i_I(M) \iso \lim_{n\to \infty} \Ext^i_A(A/I^n, M)\,,
\end{equation} 
where the isomorphism is natural, i.e., functorial. The authors of \cite{ToricCoh} extend this result to {\bf graded} rings and modules, and produce explicit finite bounds above which, at a given degree, the direct limit in (\ref{e:lc3}) is stationary. Without further ado, we present their results.

\subsection{Computing graded local cohomology}

\begin{prop}
There exist functions $f_i\!: \D\longrightarrow {\bf N}^*$, for all $i\geq 0$, such that
\[
{\mytext{Ext}}^{\,i}_S(S/B^l, S)_\delta\longrightarrow \H_B^i(S)_\delta
\]
is an isomorphism for every $\delta\in\D$ and $l\geq f_i(\delta)$. Moreover, for $P$ a finitely generated $\D$-graded $S$-module, and $F_{\bullet}$  a free resolution of $P$, with
$$F_i=\bigoplus_{\alpha\in\D}S(-\alpha)^{\oplus\beta_{i,\alpha}}\,$$
the map 
\[
{\mytext{Ext}}^{\,i}_S(S/B^l, P)_\delta\longrightarrow \H^i_B(P)_\delta
\]
is an isomorphism if for every $j\geq 0$ and every $\alpha$ such that $\oplus\beta_{j,\alpha}\neq 0$ we have
\begin{itemize}
\item  $l\geq f_i(\delta-\alpha)$, if $j=0$;
\item  $l\geq\max\{f_{i+j-1}(\delta-\alpha), f_{i+j}(\delta-\alpha)\}$, if $j>0$.
\end{itemize}
\end{prop}

Translating this into sheaf cohomology is immediate:\footnote{Both the printed and the electronic versions of \cite{ToricCoh} have typos in this statement.}

\begin{thm}
There exist functions $f_i\!: \D\longrightarrow {\N}^*$ for all $i\geq 0$ such that:
\[
{\mytext{Ext}}^{\,i+1}_S(S/B^l, S)_\delta\longrightarrow \H^i(\O_X(\delta))
\]
is an isomorphism for every $\delta\in\D$ and $l\geq f_i(\delta)$. Moreover, for $P$ a finitely generated graded $S$ module, with $F_{\bullet}$  a minimal free resolution of $P$, and
$$F_i=\bigoplus_{\alpha\in\D}S(-\alpha)^{\oplus\beta_{i,\alpha}}\,,$$
the map 
\[
{\mytext{Ext}}^{\,i+1}_S(S/B^l, P)_\delta\longrightarrow H^i_{tot}(\tilde{P})_\delta
\]
is an isomorphism if for every $j\geq 0$ and every $\alpha$ such that $\oplus\beta_{j,\alpha}\neq 0$ we have
\begin{itemize}
\item  $l\geq f_{i+1}(\delta-\alpha)$, if $j=0$;
\item  $l\geq\max\{f_{i+j}(\delta-\alpha), f_{i+j+1}(\delta-\alpha)\}$, if $j>0$.
\end{itemize}
\end{thm}

This gives a very practical and efficient way to compute local cohomology, and, as outlined above, sheaf  cohomology on toric varieties. ``All'' one needs to do is evaluate the $\Ext^{\,i}_A(A/I^n, M)$ groups and keep track of the grading. Although this is a formidable task in general for humans, machines are very good at it. We implemented such routines using the symbolic algebra program {\em Macaulay~2} written by Dan Grayson and Mike Stillman \cite{M2}.\footnote{See \cite{Eisenbud:Stillman} for a review on how {\em Macaulay~2} can be used to solve all sorts of problems in algebraic geometry.}


\end{appendix}


\begin{thebibliography}{99}

\bibitem{Klebanov-Witten}
I.~R.~Klebanov and E.~Witten,
{\em Superconformal field theory on threebranes at a Calabi-Yau  singularity},
Nucl.\ Phys.\ B {\bf 536}, 199 (1998)
[arXiv:hep-th/9807080].

\bibitem{DM}
M.~R.~Douglas and G.~W.~Moore,
  {\em D-branes, Quivers, and ALE Instantons},
  [arXiv:hep-th/9603167].

\bibitem{Gauntlett}
J.~P.~Gauntlett, D.~Martelli, J.~Sparks and D.~Waldram,
{\em Supersymmetric AdS(5) solutions of M-theory},
Class.\ Quant.\ Grav.\  {\bf 21}, 4335 (2004)
[arXiv:hep-th/0402153]. \\
%
J.~P.~Gauntlett, D.~Martelli, J.~Sparks and D.~Waldram,
{\em Sasaki-Einstein metrics on S(2) x S(3)},
[arXiv:hep-th/0403002].

\bibitem{Cvetic}
M.~Cvetic, H.~Lu, D.~N.~Page and C.~N.~Pope,
  {\em New Einstein-Sasaki spaces in five and higher dimensions},
  [arXiv:hep-th/0504225]; \\
M.~Cvetic, H.~Lu, D.~N.~Page and C.~N.~Pope,
  {\em New Einstein-Sasaki and Einstein spaces from Kerr-de Sitter},
  [arXiv:hep-th/0505223]; \\
D.~Martelli and J.~Sparks,
  {\em Toric Sasaki-Einstein metrics on $S^2 \times S^3$},
  Phys.\ Lett.\ B {\bf 621}, 208 (2005)
  [arXiv:hep-th/0505027].

\bibitem{Rudakov} 
{\em Helices and vector bundles},
Seminaire Rudakov. London Mathematical Society Lecture Note Series, 148. 
Cambridge University Press, Cambridge, 1990.

\bibitem{mayr}
P.~Mayr,
{\em Phases of supersymmetric D-branes on Kaehler manifolds and the McKay
correspondence},
JHEP {\bf 0101}, 018 (2001)
[arXiv:hep-th/0010223]. \\
A.~Tomasiello,
{\em D-branes on Calabi-Yau manifolds and helices},
JHEP {\bf 0102}, 008 (2001)
[arXiv:hep-th/0010217]. \\
S.~Govindarajan and T.~Jayaraman,
{\em D-branes, exceptional sheaves and quivers on Calabi-Yau manifolds: From
Mukai to McKay},
Nucl.\ Phys.\ B {\bf 600}, 457 (2001)
[arXiv:hep-th/0010196].

\bibitem{zaslow}
E.~Zaslow,
{\em Solitons and helices: The Search for a math physics bridge},
Commun.\ Math.\ Phys.\  {\bf 175}, 337 (1996)
[arXiv:hep-th/9408133].
\bibitem{hiv}
K.~Hori, A.~Iqbal and C.~Vafa,
{\em D-branes and mirror symmetry},
arXiv:hep-th/0005247.

\bibitem{Aspinwall:2001zq}
  P.~S.~Aspinwall,
{\em Some navigation rules for D-brane monodromy},
J.\ Math.\ Phys.\  {\bf 42}, 5534 (2001)
[arXiv:hep-th/0102198].

\bibitem{Wijnholt}
M.~Wijnholt,
  {\em Large volume perspective on branes at singularities},
  Adv.\ Theor.\ Math.\ Phys.\  {\bf 7}, 1117 (2004)
  [arXiv:hep-th/0212021].

\bibitem{unify}
F.~Cachazo, B.~Fiol, K.~A.~Intriligator, S.~Katz and C.~Vafa,
{\em A geometric unification of dualities},
Nucl.\ Phys.\ B {\bf 628}, 3 (2002)
[arXiv:hep-th/0110028].

\bibitem{HerzogWalcher}
 C.~P.~Herzog and J.~Walcher,
  {\em Dibaryons from exceptional collections},
  JHEP {\bf 0309}, 060 (2003)
  [arXiv:hep-th/0306298].

\bibitem{Karpov:Nogin}
B.~V. Karpov and D.~Y. Nogin,
\newblock {\em Three-block exceptional sets on del {P}ezzo surfaces},
\newblock Izv. Ross. Akad. Nauk Ser. Mat. {\bf 62} (No. 3)  (1998) 3--38,
  alg-geom/9703027.
  
\bibitem{Seiberg}
N.~Seiberg,
{\em Electric - magnetic duality in supersymmetric non-Abelian gauge theories},
Nucl.\ Phys.\ B {\bf 435}, 129 (1995)
[arXiv:hep-th/9411149].

\bibitem{FengHananyHe}
B.~Feng, A.~Hanany and Y.~H.~He,
  {\em Phase structure of D-brane gauge theories and toric duality},
  JHEP {\bf 0108}, 040 (2001)
  [arXiv:hep-th/0104259].

\bibitem{BeasleyPlesser}
  C.~E.~Beasley and M.~R.~Plesser,
  {\em Toric duality is Seiberg duality},
  JHEP {\bf 0112}, 001 (2001)
  [arXiv:hep-th/0109053].

\bibitem{ToricCoh}
D.~Eisenbud, M.~Mustata, and M.~Stillman,
\newblock {\em Cohomology on toric varieties and local cohomology with monomial
  supports},
\newblock J. Symbolic Comput. {\bf 29} (No. 4-5)  (2000) 583--600,
  [arXiv:math.AG/0001159].

\bibitem{HananyXpq}
  A.~Hanany, P.~Kazakopoulos and B.~Wecht,
 {\em A new infinite class of quiver gauge theories},
  [arXiv:hep-th/0503177].

\bibitem{BorisovDM}
L.~A. Borisov, L.~Chen, and G.~G. Smith,
\newblock {\em The orbifold {C}how ring of toric {D}eligne-{M}umford stacks},
\newblock J. Amer. Math. Soc. {\bf 18} (No. 1)  (2005) 193--215,
 [arXiv:math.AG/0309229].

\bibitem{Herzog1}
C.~P.~Herzog,
  {\em Seiberg duality is an exceptional mutation},
  JHEP {\bf 0408}, 064 (2004)
  [arXiv:hep-th/0405118].

\bibitem{AspinwallMelnikov}
 P.~S.~Aspinwall and I.~V.~Melnikov,
  {\em D-branes on vanishing del Pezzo surfaces},
  JHEP {\bf 0412}, 042 (2004)
  [arXiv:hep-th/0405134].

\bibitem{Fulton}
W.~Fulton, {\em Introduction to Toric Varieties}, Princeton University Press, 1993.

\bibitem{KO}
S.~A.~Kuleshov and D.~O.~Orlov,
{\em Exceptional sheaves over del Pezzo surfaces}, 
{Izv. Ross. Akad. Nauk Ser. Matem.} {\bf 58}:3 (1994)
53; 
English transl. {Russian Acad. Sci. Izv. Math.}
{\bf 44} (1995), 479.

\bibitem{Kawamata}
Y.~Kawamata,
\newblock {\em Derived Categories of Toric Varieties},
\newblock [arXiv:math.AG/0503102].

\bibitem{Herzog2}
C.~P.~Herzog,
  {\em Exceptional collections and del Pezzo gauge theories},
  JHEP {\bf 0404}, 069 (2004)
  [arXiv:hep-th/0310262].

\bibitem{Hille}
L.~Hille, 
{\em Exceptional sequences of line bundles on toric varieties},
[www.math.uni-hamburg.de/home/hille/publ.html].

\bibitem{Hananyold}
B.~Feng, A.~Hanany, Y.~H.~He and A.~M.~Uranga,
 {\em Toric duality as Seiberg duality and brane diamonds},
  JHEP {\bf 0112}, 035 (2001)
  [arXiv:hep-th/0109063].


\bibitem{Narayan}
D.~R.~Morrison, K.~Narayan and M.~R.~Plesser,
  {\em Localized tachyons in C(3)/Z(N)},
  JHEP {\bf 0408}, 047 (2004)
  [arXiv:hep-th/0406039].

\bibitem{Auraux}
D.~Auroux, L.~Katzarkov, D.~Orlov,
{\em Mirror symmetry for weighted projective planes and 
their noncommutative deformations},
[arXiv:math.AG/0404281].

\bibitem{Benvenuti}
  S.~Benvenuti, S.~Franco, A.~Hanany, D.~Martelli and J.~Sparks,
 {\em An infinite family of superconformal quiver gauge theories with
 Sasaki-Einstein duals},
  [arXiv:hep-th/0411264].

\bibitem{MartelliSparks}
  D.~Martelli and J.~Sparks,
  {\em Toric geometry, Sasaki-Einstein manifolds and a new infinite class of
  AdS/CFT duals},
  [arXiv:hep-th/0411238].

\bibitem{Brig:flop}
T.~Bridgeland,
\newblock {\em Flops and derived categories},
\newblock Invent. Math. {\bf 147} (No. 3)  (2002) 613--632, [arXiv:math.AG/0009053].

\bibitem{Kawamata:DC}
Y.~Kawamata,
\newblock {\em Equivalences of derived categories of sheaves on smooth stacks},
\newblock [arXiv:math.AG/0210439].

\bibitem{Morrison-Plesser}
D.~R.~Morrison and M.~R.~Plesser,
{\em Non-spherical horizons. I},
Adv.\ Theor.\ Math.\ Phys.\  {\bf 3}, 1 (1999)
[arXiv:hep-th/9810201].

\bibitem{Kehagias}
A.~Kehagias,
  {\em New type IIB vacua and their F-theory interpretation},
  Phys.\ Lett.\ B {\bf 435}, 337 (1998)
  [arXiv:hep-th/9805131].

\bibitem{Benvenutitoric}
  S.~Benvenuti, A.~Hanany and P.~Kazakopoulos,
 {\em The toric phases of the Y(p,q) quivers},
  [arXiv:hep-th/0412279].

\bibitem{HananyLpqr}
  S.~Franco, A.~Hanany, D.~Martelli, J.~Sparks, D.~Vegh and B.~Wecht,
  {\em Gauge theories from toric geometry and brane tilings},
 [arXiv:hep-th/0505211].

\bibitem{ItalianLpqr}
A.~Butti, D.~Forcella and A.~Zaffaroni,
  {\em The dual superconformal theory for L(p,q,r) manifolds},
 [arXiv:hep-th/0505220].
  
  \bibitem{onemoreLpqr}
  S.~Benvenuti and M.~Kruczenski,
  {\em From Sasaki-Einstein spaces to quivers via BPS geodesics: L(p,q$|$r)},
 [arXiv:hep-th/0505206].

\bibitem{Eisenbud}
D.~Eisenbud,
\newblock {\em Commutative algebra}, Graduate Texts in Mathematics~{\bf 150},
\newblock Springer-Verlag, New York, 1995.

\bibitem{Eisenbud:Stillman}
D.~Eisenbud, D.~R. Grayson, M.~Stillman, and B.~Sturmfels,
\newblock {\em Computations in Algebraic Geometry with Macaulay 2}, Algorithms
  and Computation in Mathematics~{\bf 8},
\newblock Springer-Verlag, New York, 2002.



\bibitem{Eric:DC2}
S.~Katz, T.~Pantev, and E.~Sharpe,
\newblock {\em D-branes, orbifolds, and Ext groups},
\newblock Nucl. Phys. {\bf B673} (2003) 263--300, 
[arXiv:hep-th/0212218].

\bibitem{Chen:Ruan}
W.~Chen and Y.~Ruan,
\newblock {\em A new cohomology theory of orbifold},
\newblock Comm. Math. Phys. {\bf 248} (No. 1)  (2004) 1--31.

\bibitem{Abramov:Vistoli}
D.~Abramovich, T.~Graber, and A.~Vistoli,
\newblock {\em Algebraic orbifold quantum products},
\newblock in ``Orbifolds in mathematics and physics (Madison, WI, 2001)'',
  Contemp. Math.~{\bf 310}, pages 1--24, Amer. Math. Soc., Providence, RI,
  2002.
  
\bibitem{Horj:Borisov}
L.~A. Borisov and R.~P. Horja,
\newblock {\em On the $K$-theory of smooth toric DM stacks},
\newblock [arXiv:math.AG/0503277].

\bibitem{Hartshorne:}
R.~Hartshorne,
\newblock {\em Algebraic geometry},
\newblock Springer-Verlag, New York, 1977,
\newblock Graduate Texts in Mathematics, No. 52.

\bibitem{Cox:HoloQout}
D.~A. Cox,
\newblock {\em The homogeneous coordinate ring of a toric variety},
\newblock J. Algebraic Geom. {\bf 4} (No. 1)  (1995) 17--50, 
[arXiv:alg-geom/9210008].

\bibitem{en:fracC3}
R.~L. Karp,
\newblock {\em On the $\C^n/\Z_m$ fractional branes},
\newblock hep-th/0602165.

\bibitem{en:fracC2}
R.~L. Karp,
\newblock {\em $\C^2/\Z_n$ Fractional branes and Monodromy},
\newblock hep-th/0510047.

\bibitem{Cox:Katz}
D.~A. Cox and S.~Katz,
\newblock {\em Mirror symmetry and algebraic geometry}, Mathematical Surveys
  and Monographs~{\bf 68},
\newblock American Mathematical Society, Providence, RI, 1999.

\bibitem{Knutson:Sharpe}
A.~Knutson and E.~R. Sharpe,
\newblock {\em Sheaves on toric varieties for physics},
\newblock Adv. Theor. Math. Phys. {\bf 2} (1998) 865--948, 
[arXiv:hep-th/9711036].

\bibitem{Klyachko}
A.~A. Klyachko,
\newblock {\em Equivariant bundles over toric varieties},
\newblock Izv. Akad. Nauk SSSR Ser. Mat. {\bf 53} (No. 5)  (1989) 1001--1039,
  1135.

\bibitem{Witten:GLSM}
E.~Witten,
\newblock {\em Phases of N = 2 theories in two dimensions},
\newblock Nucl. Phys. {\bf B403} (1993) 159--222, 
[arXiv:hep-th/9301042].

\bibitem{LocalCohomology}
M.~P. Brodmann and R.~Y. Sharp,
\newblock {\em Local cohomology: an algebraic introduction with geometric
  applications}, Cambridge Studies in Advanced Mathematics~{\bf 60},
\newblock Cambridge University Press, Cambridge, 1998.

\bibitem{Hartshorne:Local}
R.~Hartshorne,
\newblock {\em Local cohomology}, A seminar given by A. Grothendieck, Harvard
  University~{\bf 1961},
\newblock Springer-Verlag, Berlin, 1967.

\bibitem{M2}
D.~R. Grayson and M.~E. Stillman,
\newblock {\em Macaulay 2, a software system for research in algebraic
  geometry},
\newblock Available at http://www.math.uiuc.edu/Macaulay2.


\bibitem{Distler:1996tj}
J.~Distler, B.~R. Greene, and D.~R. Morrison,
\newblock {\em Resolving singularities in (0,2) models},
\newblock Nucl. Phys. {\bf B481} (1996) 289--312, 
[arXiv:hep-th/9605222].

\bibitem{Aspinwall-Katz}
P.~S.~Aspinwall and S.~Katz,
  {\em Computation of superpotentials for D-Branes},
  [arXiv:hep-th/0412209.

\bibitem{Aspinwall-Fidkowski}
P.~S.~Aspinwall and L.~M.~Fidkowski,
  {\em Superpotentials for quiver gauge theories},
  [arXiv:hep-th/0506041.

\end{thebibliography}

\end{document}